\documentclass[fleqn,usenatbib]{mnras}

\usepackage{newtxtext,newtxmath}

\usepackage[T1]{fontenc}
\usepackage{ae,aecompl}
\usepackage[dvipsnames]{xcolor}
\definecolor{pink}{rgb}{0.98, 0.38, 0.5}

\usepackage{graphicx}	
\usepackage{amsmath}	
\usepackage{amssymb}	
\usepackage{ tipa }
\usepackage{xtab,afterpage}
\usepackage{longtable}
\usepackage{cleveref}
\crefformat{section}{\S#2#1#3}

\def\teff{$T_{\rm eff}$}
\newcommand{\logg}{\mbox{$\log g$}}
\newcommand{\vmicro}{$\rm v_{micro}$}

\newcommand{\logeps}{log$(\epsilon)$}

\title[abundance patterns of accreted stars]{The detailed chemical abundance patterns of accreted halo stars from the optical to infrared}

\author[A. Carrillo et al.]{
Andreia Carrillo$^{1,2,3}$\thanks{E-mail: andreiac@utexas.edu},
Keith Hawkins$^{1}$,
Paula Jofr\'e$^{4}$,
Danielle de Brito Silva$^{4}$,
Payel Das$^{5}$,
\newauthor and Madeline Lucey$^{1}$
\\
$^{1}$Department of Astronomy, University of Texas at Austin, 2515 Speedway, Stop C1400, Austin, TX 78712-1205, USA\\
$^{2}$Institute for Computational Cosmology, Department of Physics, Durham University, Durham DH1 3LE, UK\\
$^{3}$Large Synoptic Survey Telescope Corporation Data Science Fellow\\
$^{4}$N\'ucleo de Astronom\'ia, Facultad de Ingenier\'ia y Ciencias, Universidad Diego Portales, Ejercito 441, Santiago, Chile\\
$^{5}$Department of Physics, University of Surrey, Guildford GU2 7XH, UK\\
}

\date{Accepted XXX. Received YYY; in original form ZZZ}

\pubyear{2019}

\begin{document}
\label{firstpage}
\pagerange{\pageref{firstpage}--\pageref{lastpage}}
\maketitle

\begin{abstract}
Understanding the assembly of our Galaxy requires us to also characterize the systems that helped build it. In this work, we accomplish this by exploring the chemistry of accreted halo stars from the Gaia-Enceladus/Gaia-Sausage (GES) selected in the infrared from the Apache Point Observatory Galactic Evolution Experiment (APOGEE) Data Release 16. We use high resolution optical spectra for 62 GES stars to measure abundances in 20 elements spanning the $\alpha$, Fe-peak, light, odd-Z, and notably, the neutron-capture groups of elements to understand their trends in the context of and in contrast to the Milky Way and other stellar populations. Using these derived abundances we find that the optical and the infrared abundances agree to within 0.15 dex except for O, Co, Na, Cu, and Ce. These stars have enhanced neutron-capture abundance trends compared to the Milky Way, and their [Eu/Mg] and neutron-capture abundance ratios (e.g., [Y/Eu], [Ba/Eu], [Zr/Ba], [La/Ba], and [Nd/Ba]) point to r-process enhancement and a delay in s-process enrichment. Their [$\alpha$/Fe] trend is lower than the Milky Way trend for [Fe/H]$>$-1.5 dex, similar to previous studies of GES stars and consistent with the picture that these stars formed in a system with a lower rate of star formation. This is further supported by their depleted abundances in Ni, Na, and Cu abundances, again, similar to previous studies of low-$\alpha$ stars with accreted origins.  
\end{abstract}




\section{Introduction}
The hierarchical model posits two phases to the formation of larger galactic systems: a first phase dominated by mergers between similar-sized galaxies, and a second phase dominated by accretion of smaller systems. 
We are especially able to study such interactions in the Milky Way where we can resolve the individual stars. There are streams of stars in the halo,  remnants of smaller systems, i.e. galaxies or globular clusters, being tidally disrupted by our Galaxy's potential (e.g., \citealt{ibata94, lyndenbell95,belokurov06,bonaca12,shipp18}). However, this is only evidence for recent interactions with the Milky Way---there is little spatially-distinct evidence for past merger events, as these accreted stars would have been phase-mixed in the Galaxy over time. These ex-situ stars no longer cluster spatially but they may be identifiable through their kinematics and chemical tagging \citep{freeman02}. Through doing this, we are able to recover one of Milky Way's most significant merger event \citep{helmi18}.


To study the systems that the Milky Way has accreted through its lifetime, we search for old, metal-poor stars in the halo. Low-metallicity, kinematically hot, halo stars in the Solar neighborhood have been shown to have two chemically-distinct stellar populations -- one with high [$\alpha$/Fe]\footnote{The bracket notation denotes differential analysis with the Sun i.e. [X/H] = log($N_X$/$N_{H}$)$_{\star}$ - log($N_X$/$N_{H}$)$_{\odot}$ = log$A_{X,\star}$ - log$A_{X,\odot}$  where log$N_H$ is 12.00. Alternatively, abundance is reported with respect to the Solar Fe abundance, i.e. [X/Fe] = [X/H] - [Fe/H].  Additionally, [$\alpha$/Fe] is the average abundance of $\alpha$ elements (e.g.  O, Mg, Ca, Si) with respect to Solar values. } abundance and one with low [$\alpha$/Fe] abundance \citep{nissen97,fulbright02,Nissen10}. The kinematics, orbital parameters, and ages of these stars show that they may have formed through very different scenarios \citep{schuster12}. The low-$\alpha$ stars (at metallicities, [Fe/H] $\geq$-1.5 dex) are postulated to have come from a dwarf galaxy with a lower star formation rate and different chemical evolution than the Milky Way progenitor. On the other hand, the high-$\alpha$ stars at the same [Fe/H] are thought to be pre-existing in-situ disk material, dispersed to halo kinematics in the infancy of our Galaxy, because of a significant merger event \citep{zolotov09,zolotov10,purcell10}. 

Investigation on the origin of these  low-$\alpha$ stars 
has been further advanced with the advent of the European Space Agency satellite \textit{Gaia} \citep{gaia16} whose second data release (DR2, \citealt{gaia18}) revolutionized the 3D map of our Galaxy. Having observed 1.7 billion stars, the survey provides astrometric, photometric, and for a subset, radial velocity (RV) information. This has especially opened the doors to reconstructing accretion events in the Milky Way as they would leave considerable substructure in phase-space \citep{johnston96}. \citet{helmi18} used \textit{Gaia} DR2 to kinematically select accreted stars that once belonged to the system that the authors dubbed Gaia-Enceladus. Gaia-Enceladus stars cover almost the full sky but systematically show slight retrograde motion, which is seen as evidence that they once belonged to a singular system separate from the Milky Way. \citet{belokurov18} found the same population of stars using the Sloan Digital Sky Survey (SDSS)-\textit{Gaia} catalogue \citep{ahn12,gaia16dr1} that contains proper motions (from \textit{Gaia} DR1) and [Fe/H]. They found more radial orbits for stars in the inner halo and a change in the halo velocity ellipsoid, $\beta$, as a function of [Fe/H] and height from the disc plane. These results are best supported by a formation scenario where the Milky Way accreted a system with $M_{vir} > 10^{10} M_{\odot}$ during the epoch of disc formation, 8 to 11 Gyr ago, later called Gaia-Sausage. In this work, we refer to this system as the Gaia-Enceladus-Sausage (GES) galaxy.

Detailed chemical abundance studies using large spectroscopic surveys, like the high-resolution infrared (IR) survey Apache Point Observatory (APO) Galactic Evolution Experiment (APOGEE, \citealt{eisenstein11}), have further enlightened the origin of these accreted stars and for a much larger sample. \citet{hawkins15} used APOGEE DR12 \citep{holtzman15} to show that the accreted inner halo stars are distinct in the [$\alpha$/Fe], [Mn/Fe], and [Al/Fe] vs [Fe/H] planes to name a few, hinting at a slower chemical evolution and star formation history. Combinations of these chemical abundances are therefore promising in disentangling the stars that formed inside and outside of the Milky Way.  \citet{hayes18}, in a similar vein, used the $\alpha$ element Mg to separate low-metallicity stars in APOGEE DR 13 \citep{albareti17} into a low-Mg and high-Mg population. They find that the two groups are also distinct from each other in other chemical abundance planes and that the low-Mg population has halo-like kinematics and are most likely accreted. Combining the power of chemistry from APOGEE DR14 and kinematics from Gaia DR2, \citet{mackereth19} found that the canonical halo stars defined in the [Mg/Fe] vs [Fe/H] plane can be divided into a low eccentricity, low-\textit{e}, and high eccentricity, high-\textit{e}, group. They find that the high-\textit{e} stars show lower Mg, Al, and Ni abundances compared to the low-\textit{e} group, reaffirming the distinguishing elements suggested from \citet{hawkins15}. To explain the existence of these high-\textit{e} stars, \citet{mackereth19} supplemented their study with comparisons to the EAGLE simulation and found that the high \textit{e} (i.e. 0.85) can be explained by a merger that happened at z$\lesssim$1.5 with stellar mass between $\rm 10^{8.5} < M_{\star} < 10^{9}$ $M_{\odot}$. 

Using the combination of elements presented by \citet{hawkins15} to purely chemically select stars accreted onto the Milky Way, \citet{das20} further showed that the combination of APOGEE (DR14) and Gaia DR2 is a powerful tool in disentangling ex-situ vs in-situ stars, and characterizing their kinematics and ages. The authors find a ``blob" of 856 likely accreted stars in the [Mg/Mn] vs [Al/Fe] plane and with a Bayesian isochrone pipeline, derived ages for these blob stars that range between 8 and 13 Gyr. With a single progenitor scenario and from dynamical arguments, these blob stars likely belonged to a system with a total mass of $\rm \sim 10^{11} M_{\odot}$, where the star formation proceeded for 5 Gyr until it was cannibalized by the Milky Way 8 Gyr ago.

Combining the power of large Milky Way surveys has enabled the community to further characterize the GES system but to fully understand this progenitor's chemistry and its scatter in abundances, we need access to many more elements. APOGEE DR16 \citep{jonsson20} provides abundances for $>$20 elements, including Cerium (Ce) and Neodymium (Nd), both neutron-capture elements. Neutron-capture elements are thought to play a bigger role in the chemical evolution of lower mass systems. Simulations show that slow (s)-process neutron capture element abundance can can be used to distinguish the origin of the halo, especially at lower metallicities \citep{spitoni16}. \citet{nissen11} and \citet{fishlock17} specifically derived predominant s-process element abundances for low-$\alpha$ stars in the inner halo and concluded that low-metallicity, low-mass asymptotic giant branch (AGB) stars dispersed metals in the progenitor with a time-delay. More recent studies (e.g. \citealt{matsuno20}, \citealt{matsuno21},\citealt{aguado20}) also find r-process enhancement in GES stars. Most of the studies aforementioned have samples of 10-30 low-$\alpha$ stars, but as been shown with more recent studies using large surveys (e.g. \citealt{belokurov18,helmi18,hayes18,mackereth19,bird19,lancaster19}), it is possible to select accreted halo stars en masse. However, most neutron-capture elements like Barium (Ba), Lanthanum (La), Neodymium (Nd), Yttrium (Y), Zirconium (Zr), and (the predominantly r-process element) Europium (Eu), are harder to measure, if not impossible, in the IR compared to the optical. Some neutron-capture elements are available in other optical spectroscopic surveys (e.g. GALAH, \citealt{buder18}) and have indeed been used in the more recent GES literature \citep{matsuno21,aguado20}. However, a sample selected from giants in IR data (such as APOGEE) probes stars at larger distances.

Therefore, in this work, we present a complementary study of the GES system: motivated by \citet{das20}, we take advantage of the power of chemical tagging and select a sample of accreted stars based solely on their chemistry in the IR and derive their detailed chemical abundances in the $\alpha$, Fe-peak, light, odd-Z, and neutron capture elements in the optical. 
We aim to showcase the 
elemental abundance trends of the GES system particularly in the neutron-capture elements, to understand its enrichment and chemistry in the context of the satellite population of the Milky Way. We also seek to compare the detailed chemical abundances between the optical and IR, and quantify their systematic differences for the same set of stars at such low metallicities. This paper is outlined as followed: Section \ref{sec:data} describes our target selection of accreted halo stars that form a blob in the [Mg/Mn] vs [Al/Fe] plane and outline our observations. We refer to our sample as blob stars. Section \ref{sec:stellarparams} outlines how we derived the stellar parameters and the abundances. Section \ref{sec:OpticalvsIR} compares the stellar parameters and abundances between what we derived vs their APOGEE values. Section \ref{sec:abundances} discusses the [X/Fe] vs [Fe/H] for our sample of stars in contrast with different stellar populations. Section \ref{sec:discussion} considers the implications of our sample's abundance trends on its progenitor's chemical evolution and its context within the rest of the Milky Way satellite population. Lastly, Section \ref{sec:summary} summarizes the main results from this study and avenues for future work.  


\section{Data}
\label{sec:data}
\subsection{Accreted Stars Selection}
\label{sec:selection}

\begin{figure}
\includegraphics[width=0.49\textwidth]{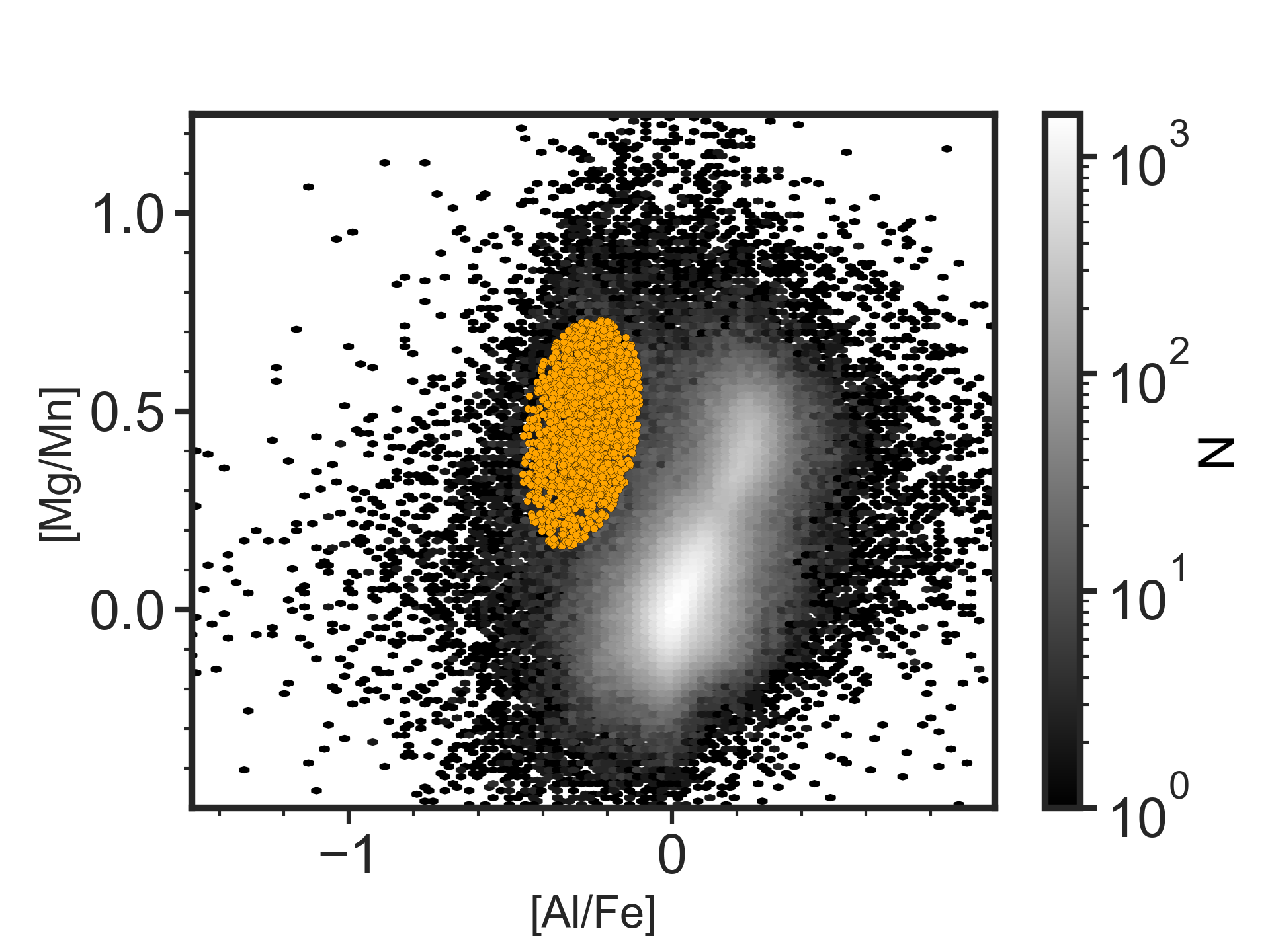}
\caption{\textbf{\textit{Chemically selecting stars with likely accretion origin}}. [Mg/Mn] vs [Al/Fe] that shows the 2,258 accreted stars as a ``blob" in this plane (orange), in contrast with the rest of the stars from APOGEE DR16 (shown as logarithm of the number count in gray). The colorbar indicates the number of stars per hexbin in the APOGEE DR16 data.} 
\label{fig:blob}
\end{figure}

We employ a pure chemical selection of accreted halo stars using APOGEE DR16 \citep{ahumada19,jonsson20}. APOGEE has moderate resolution (R $\sim$ 22,500) \textit{H}-band spectra (1.5-1.7 $\mu$m) for 473,307 stars in the 16th data release (DR16), taken both in the Northern hemisphere with the Sloan Foundation 2.5-m Telescope at APO (APOGEE-2N) and in the Southern hemisphere with the du Pont 2.5-m Telescope at Las Campanas Observatory (LCO).  

The stellar parameters and abundances were derived through the APOGEE Stellar Parameters and Chemical Abundances Pipeline (ASPCAP; \citealt{garciaperez16}). ASPCAP determines the best-matching synthetic spectra that has known stellar parameters to the observed spectra. A new set of synthetic grids have been utilized for this analysis using MARCS model atmospheres \citep{gustafsson08} that cover a wide range in surface gravity (log $g$), effective temperature (\teff), microturbulence ($\rm v_{micro}$), [Fe/H], and abundance parameters [$\alpha$/M], [C/M], and [N/M]. APOGEE contains chemical abundances, [X/Fe], based on the molecular lines of  C, N, O, and Si, and atomic lines of C, Na, Mg, Al, Si, P, S, K, Ca, Ti I, Ti II, V, Cr, Mn, Fe, Co, Ni, Cu, Ge, Rb, Ce, Nd, and Yb.

Using APOGEE DR12 data, \citet{hawkins15} introduced a new combination of chemical abundances i.e. [C+N/Fe], [Al/Fe], and [Mg/Mn] that could discern stars belonging to different Galactic components, independent of their kinematics. \citet{mackereth19} reaffirmed this later, and showed that not only [Mg/Fe] but also [Al/Fe] and [Ni/Fe] separate stars that have likely ex-situ origins as shown by their high-$e$. These elements provide more ways to separate the in-situ and ex-situ Milky Way stars, in addition to the canonically-used  [$\alpha$/Fe]. In this study, we use APOGEE DR16 to select targets with potential accreted origin in the [Mg/Mn] vs [Al/Fe] plane.


While APOGEE DR16 contains information for several elements, including neutron capture elements Ce and Nd (\citealt{jonsson20}), these are typically difficult to measure for low metallicity stars. We therefore obtain high-resolution optical spectra for better derivation of neutron-capture element abundances. 



We followed the method from \citet{das20} in selecting the likely accreted stars in the [Mg/Mn] vs [Al/Fe] plane. Selecting in chemistry is particularly advantageous because the stellar atmospheric abundances are essentially preserved throughout a star's lifetime except for some light elements, such as Li, C, and N due to dredge-up processes. This is especially promising as \citet{das20} showed that using this combination of abundances, their sample of accreted stars have older ages (see their Figure 7) and kinematics (see their Figure 8) similar to accreted stars in the literature (e.g., \citealt{belokurov18}). Whether selecting dynamically or chemically gives a cleaner or different sample of accreted stars is worth exploring and has been the subject of the work from \citet{buder22}. Using GALAH DR3 data and the combination of [Na/Fe] and [Mg/Mn], the authors found that their GES sample from a chemical selection overlaps with 29\% of a pure dynamically-selected sample, as suggested by \cite{feuillet21}. Likewise, there is a mismatch for their dynamically-selected GES stars with 72\% of the sample having [Na/Fe] abundance above the 84th percentile of the chemical selection.

In this work, we applied Gaussian Mixture Modeling (GMM)~\footnote{https://scikit-learn.org/stable/modules/mixture.html} in the [Mg/Mn] vs [Al/Fe] plane. GMM is a probablistic model that assumes that the data is drawn from a mixture of a finite number of Gaussian distributions. However, determining the number of Gaussian distributions is non-trivial, especially for data that do not have distinctly separate distributions. To determine the optimal number of Gaussian components, we compared GMMs with one to 20 components and find that the optimal number is 17 using Bayesian Information Criterion (BIC). As it is a probablistic model, each star has a probability to be part of one of the 17 components. We determined the cutoff probability to be $>$0.7 in assigning a star to a certain Gaussian component. There are a total of 2,258 sources in the red giant branch associated with the accreted component, where the stars form a ``blob" in the [Mg/Mn] vs [Al/Fe] plane, as shown in Figure \ref{fig:blob}. We similarly refer to our sample as ``blob stars".  In contrast, \citet{das20} used APOGEE DR14, had 14 components from the GMM, and found 856 stars associated to their blob sample with 7\% contamination where some stars picked out by the GMM have Milky Way disk kinematics and [Fe/H] (see their Figure 5). In order to account for this contamination in our sample and pick purely accreted halo material, we selected only lower-metallicity stars with [Fe/H] $<$-0.8 dex. We also note that one of the stars in our blob sample, 2M14534136+4304352, was part of the blob in APOGEE DR14 \citep{das20}, though not in APOGEE DR16. We include it nonetheless. 


\begin{figure*}
\centering
\includegraphics[width=\textwidth]{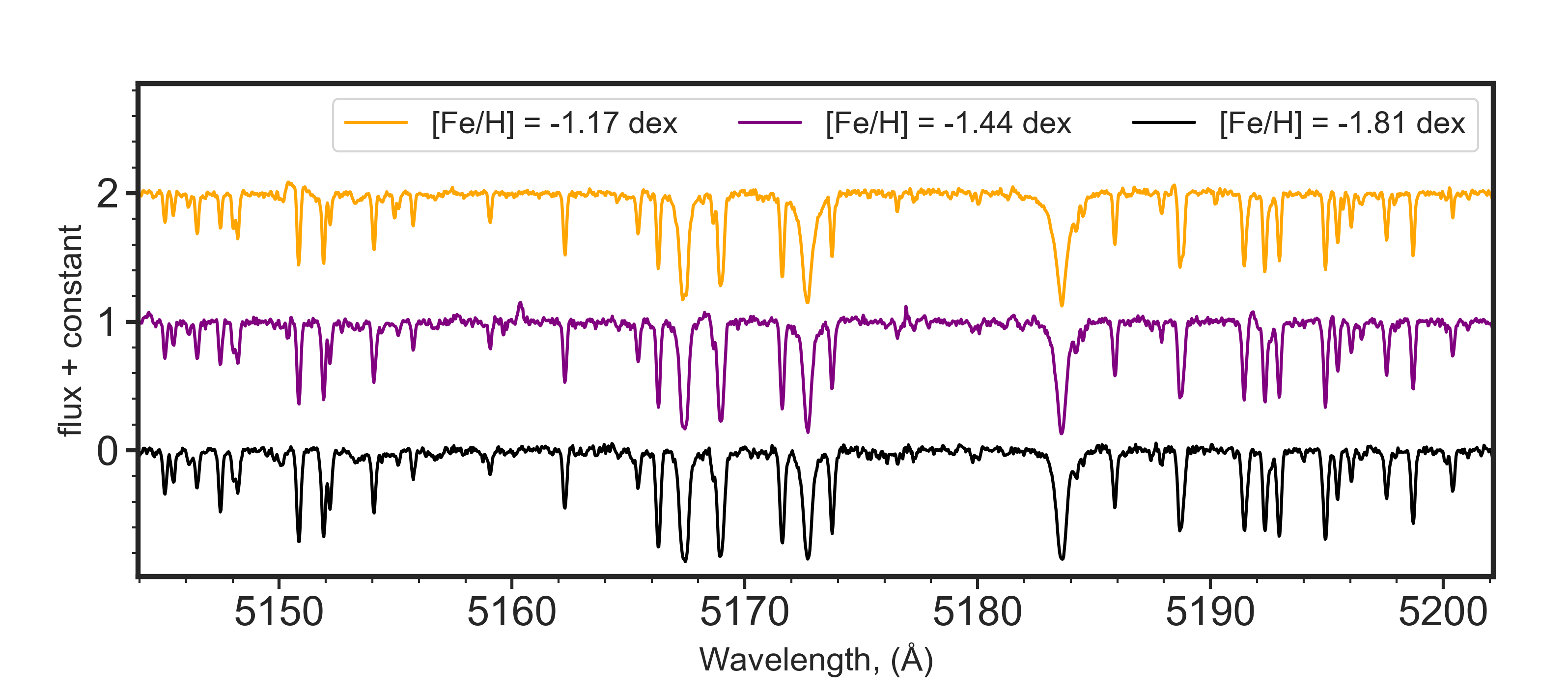}
\caption{\textit{\textbf{Continuum normalized optical spectra for three blob stars.}} They include 2M15561756+4703281 with [Fe/H] = -1.17 dex (orange), 2M10313588+3010198 with [Fe/H] = -1.44 dex (purple), and 2M12071560+4622126 with [Fe/H] = -1.81 dex (orange). }
\label{fig:blob_spectra}
\end{figure*}

\begin{table*}
\begin{tabular}{cccccccc}
\hline \hline
APOGEE ID & RA (deg) & Dec (deg) & Gaia DR2 source ID & G (mag) & RV ($\rm km s^{-1}$) & $\rm \sigma_{RV}$ ($\rm km s^{-1}$) & SNR ($\rm pixel^{-1}$)\\
\hline
2M09121759+4408563 & 138.07 & 44.15 & 817319945578821504 & 11.43 & -268.14 & 0.27 & 86.67 \\
2M11115726+4551087 & 167.99 & 45.85 & 788225596597420160 & 10.76 & -51.21 & 0.17 & 72.86 \\
2M12071560+4622126 & 181.82 & 46.37 & 1539803504272415232 & 10.88 & -179.02 & 0.20 & 84.85 \\
2M13581572+2602122 & 209.57 & 26.04 & 1450722893255972992 & 11.18 & -85.31 & 0.31 & 62.80\\
2M09381836+3706176 & 144.58 & 37.10 & 799015207281513600 & 10.63 & 187.96 & 0.13 & 85.06\\
2M11482205-0030318 & 177.09 & -0.51 & 3794814922703318272 & 11.09 & 40.64 & 0.28 & 66.81\\
2M10013420+4345558 & 150.39 & 43.77 & 808093771712305664 & 11.36 & -107.64 & 0.17 & 82.65 \\
2M13015242+2911180 & 195.47 & 29.19 & 1464266230811360256 & 11.07 & -207.34 & 0.25 & 75.82\\
2M14181562+4651580 & 214.57 & 46.87 & 1506860143038459520 & 11.23 & -46.52 & 0.28 & 95.55\\
2M15410952+3014067 & 235.29 & 30.24 & 1273071638461821952 & 10.75 & -97.31 & 0.17 & 66.98\\
\hline\hline
\end{tabular}
\caption{Properties for 10 out of 62 blob targets including the APOGEE ID, RA and Dec of the star from APOGEE, Gaia DR2 source ID, Gaia G-band magnitude, radial velocity and error corrected for barycentric motion derived from the optical spectra, and SNR. Full version is provided online. }
\label{tab:targets}
\end{table*}

\subsection{High Resolution Optical Spectroscopy of Accreted Stars}
\label{sec:observation}

We selected stars with $G$ $<$ 12 mag and obtained high-resolution optical spectra with the Tull Echelle Spectrograph \citep{tull95} at the 2.7-m Harlan J. Smith telescope at McDonald Observatory. 
We used the Coud\`e spectrograph in the TS23 mode, utilizing slit\#5, with width 1.79" and full width half maximum (FWHM) 2.78 pixels. This achieves R $\sim$40,000 over a wavelength range of 3400 -- 9000~~\AA. Typical exposure times range from 1200 to 1800 seconds and the number of exposures vary from one to four, depending on the magnitude of the target and the weather conditions, in order to reach moderate signal-to-noise ratio (SNR) of at least 60~pixel$^{-1}$ 
and remove cosmic rays for the fainter targets. For every night, we observed a solar port spectrum (to ensure that our setup does not change night to night), 
biases, flats, Thorium-Argon (ThAr) comparison frames for wavelength calibration, telluric and RV standards, and a twilight spectrum to derive Solar abundances for use in our analysis. 
In addition to the science targets, we also observed telluric and RV standards. 

We performed standard reduction (i.e. bias subtraction, flat-fielding, wavelength calibration) using IRAF\footnote{IRAF is distributed by the National Optical Astronomy Observatories, which is
operated by the Association of Universities for Research in Astronomy, Inc.,
under contract to the National Science Foundation.} and pyraf for the data taken at McDonald Observatory. Each spectra has $\sim$60 orders, with greater gaps between orders towards the redder part of the spectrum. We masked bad pixels as well as instrumental artifacts and subtracted scattered light. We then normalized the spectra from each order using a fifth-degree polynomial fit to the continuum, and stitched the orders to obtain the full spectra. 

We also obtained high resolution optical spectra in the Southern hemisphere  
with the Magellan Inamori Kyocera Echelle (MIKE, \citealt{berstein03}) spectrograph at the Magellan Telescopes at Las Campanas Observatory (LCO). 
We observed with the MIKE blue setup (4100-5000~~\AA) and red setup (4900-9000~~\AA) with R $\sim$ 28,000 and 22,000 respectively. 
Biases, flats, ThAr arc lamps, RV and telluric standards, and science frames were taken, with exposure times ranging between 400 and 900 seconds for the science targets. 

The MIKE  spectra were reduced using CarPy  \citep{kelson00,kelson03}. CarPy is written in python and generates science ready 2D spectra considering flats and biases taken with the same setups. For a proper wavelength solution,  ThAr arcs were taken in between observations, and CarPy considers the arc that was most recently taken from the observation. CarPy also automatically co-adds spectra of multiple exposures of the same object.   We then stitched the extracted 2D blue and red spectra using IRAF scombine to have a final 1D spectrum per star.  

We determined and corrected for RV with iSpec \citep{blancocuaresma14,blancocuaresma19} using the cross-correlation of our spectra with that of Arcturus. 
For stars that we observed at McDonald with multiple 
exposures, we co-added their spectra to increase their SNR. In the end, we have reduced, extracted, wavelength-corrected, RV-corrected, and co-added spectra for 62 blob stars. We list the observational properties of some stars from our sample in Table \ref{tab:targets} and show sample spectra in Figure \ref{fig:blob_spectra}. 

\section{Stellar Parameter and Abundance Determination}
\label{sec:stellarparams}

In this section, we outline our derivation of atmospheric parameters for the blob stars: \teff~through the infrared flux method (IRFM, \citealt{blackwell80},\citealt{ramirez05},\citealt{gonzalezhernandez09}), \logg~from isochrone fitting (e.g. \citealt{yong13, roederer14}), and \vmicro~and [Fe/H] using the Brussels Automatic Code for Characterizing High accUracy Spectra (BACCHUS, \citealt{masseron16}) code. Using these stellar parameters, we proceeded to obtain detailed chemical abundances from the high-resolution optical spectra using the BACCHUS code. 


We adopted Equation 10 from \citet{gonzalezhernandez09}, which quantifies an empirical relation to derive the \teff~ from [Fe/H] and colour. For the input to the IRFM, we used the [Fe/H] from APOGEE DR16, and the $J$ and $K_S$ mags from the Two Micron All-Sky Survey (2MASS, \citealt{skrutskie06}), which were corrected for reddening using the \texttt{dustmaps} package \citep{green18} and its built in modules for 2D \citep{planck14} or 3D corrections \citep{green19}, where available. We accounted for the appropriate coefficients for giant stars using $J-K_S$ colour from Table 5 in \citet{gonzalezhernandez09}. We derived the \logg~through isochrone fitting using the Yonsei-Yale ($Y^2$) isochrones \citep{demarque04} interpolator, that requires the following inputs: $V$-band and European Southern Observatory (ESO) $K$-band magnitudes (among other filter combinations), age, [Fe/H], [$\alpha$/Fe]. We used the [Fe/H] and [$\alpha$/Fe] from APOGEE DR16, converted the \textit{Gaia} $G$ magnitude to Johnson-Cousins $V$ using the transformation from the \textit{Gaia} Data Release Documentation \footnote{\url{https://gea.esac.esa.int/archive/documentation/GDR2/Data_processing/chap_cu5pho/sec_cu5pho_calibr/ssec_cu5pho_PhotTransf.html}}, and converted the 2MASS $K_S$-band magnitude to ESO $K$ \footnote{\url{https://www.astro.caltech.edu/~jmc/2mass/v3/transformations/}} to get an isochrone with the same properties as the star of interest. We interpolate over the isochrone to get the associated \logg~using the $M_V$ vs $V-K$ colour-magnitude diagram (CMD). To obtain the absolute magnitude $M_V$ from apparent magnitude $V$, we used the distance estimates from \citet{bailerjones21} instead of simply inverting the \textit{Gaia} parallaxes. Lastly, in creating the isochrones, we assumed a fixed age of 10 Gyr and note that modifying the age by 2 Gyr (i.e., 8 Gyr and 12 Gyr), changes the \logg~by only 0.04 dex. 

To obtain the remaining stellar atmospheric parameters (i.e., [Fe/H] and broadening from rotation and the instrument) we used the \textit{param} module in BACCHUS. BACCHUS uses line by line analysis to derive the stellar parameters and abundances. 
It utilizes the fifth version of the atomic line list from Gaia-ESO \citep{heiter2021atomic} that includes hyperfine structure (HFS) for Sc I,
V I Mn I, Co I, Cu I, Ba II, Eu II, La II, Nd II,
Sm II. Additionally, the line list also includes molecular lines for CH \citep{masseron14}, CN, NH, OH, MgH, and C2 (T. Masseron, private communication), SiH (Kurucz linelist \footnote{\url{ http://kurucz.harvard.edu/linelists/linesmol/}}), and TiO, ZrO, FeH, and CaH (B. Plez, private communication). BACCHUS uses MARCS model atmosphere grid \citep{gustafsson08} and synthesizes model spectra using TURBOSPECTRUM \citep{alvarez98,plez12}. For more details on how BACCHUS derives stellar parameters, we refer the reader to Section 4 of \citet{lucey19} and Section 3 of \citet{hawkins20}.


We employ BACCHUS and its \textit{eqw} routine to determine the equivalent widths (EW) of Fe lines 
for our sample, 
which are later used for calculating convolution, $\rm v_{micro}$ and [Fe/H]. The EWs are calculated by integrating synthesized spectra over some previously determined spectral window \citep{masseron16}. The convolution, which in BACCHUS is the broadening as a result of the amalgamation of instrumental resolution, \textit{v}sin\textit{i}, and macroturbulence,  is  determined from ensuring the Fe abundance computed from a line's core and from its EW are similar. \vmicro~is computed by requiring no correlation between Fe abundance and the reduced equivalent width (EW/wavelength). [Fe/H] convergence is determined through $\chi^{2}$ minimization between the model spectra and the observed spectra in the defined spectral window.  



With the \teff, \logg, \vmicro, and [Fe/H] determined, we then calculated the abundances for each element, X, using the \textit{abund} routine in BACCHUS. This creates synthetic spectra given the stellar parameters, at different values of log$A_X$. 

The abundance for each line of every element is determined by $\chi^2$ minimization between the synthesized spectra and the observed spectra in the defined spectral window. The final adopted atmospheric abundance is derived by computing the median [X/Fe] value for the lines of species X that BACCHUS accepts through its automatic decision tree (see \citealt{hawkins16b} for details). Generally, the solution for a line is flagged and/or rejected if it is an upper limit, an extrapolation, or saturated (i.e. not in the linear part of the curve of growth with core depth $\approx$ 0.2 in normalized flux). The internal uncertainty is computed as the standard error of the mean. For elements with only one line vetted as good, we assumed a 0.10 dex error.  

We also used BACCHUS to derive solar atmospheric parameters and chemical abundances from our observed twilight spectra \footnote{\teff=5715 K, \logg=4.41 dex, [Fe/H]=0 dex, and \vmicro=0.89 km/s}, which we then adopted for calculating the [X/Fe] for the program stars. We took this approach to further reduce the systematic offsets introduced by our chemical abundance determination. A key difference, however, is that we employed ionization-excitation balance to derive solar \teff~and~\logg, as these could be more reliably derived from solar spectra, than from our sample of low-metallicity, low-surface gravity stars. We note that our derived abundance, \logeps, values are comparable to values in the literature (e.g. \citealt{asplund05,asplund09},\citealt{guo17},\citealt{hawkins20}). The tabulated comparison is listed in Table \ref{tab:solabund}.

\begin{table}
\center
\begin{tabular}{cccccc}
\hline \hline
Element & log(X) & $\rm \sigma_{log(X)}$ & A05 & A09 \\
\hline
O & 8.72 & 0.09 &8.66 & 8.69 \\
Na & 6.26 & 0.04 &6.17 & 6.24 \\
Mg & 7.54 & 0.05 &7.53 & 7.60 \\
Si & 7.48 & 0.03 &7.51 & 7.51 \\
Ca & 6.35 & 0.02 &6.31 & 6.34 \\
Sc & 3.21 & 0.03 &3.05 & 3.15 \\
V & 3.82 & 0.02 &4.00 & 3.93 \\
Cr & 5.54 & 0.02 &5.64 & 5.64 \\
Mn & 5.26 & 0.04 &5.39 & 5.43 \\
Co & 4.75 & 0.04 &4.92 & 4.99 \\
Ni & 6.19 & 0.04 &6.23 & 6.22 \\
Fe & 7.45 & 0.01 &7.45 & 7.50 \\
Cu & 4.00 & 0.06 &4.21 & 4.19 \\
Zn & 4.46 & 0.07 &4.60 & 4.56 \\
Y & 1.96 & 0.05 &2.21 & 2.21 \\
Zr & 2.58 & 0.06 & 2.59 & 2.58 \\
Ba & 2.24 & 0.07 &2.17 & 2.18 \\
La & 1.13 & 0.06 &1.13 & 1.10 \\
Ce & 1.36 & 0.05 &1.58 & 1.58 \\
Nd & 1.22 & 0.06 &1.45 & 1.42 \\
Eu & 0.23 & 0.09 &0.52 & 0.52 \\
\hline \hline
\end{tabular}
\caption{Independent, BACCHUS-derived solar abundances from twilight spectrum taken at McDonald Observatory. We use these solar abundances in calculating the [X/Fe] of the stars in this study. We also show other reference studies for solar abundances, i.e. \citet{asplund05}, A05, and \citet{asplund09}, A09. }
\label{tab:solabund}
\end{table}

\begin{table}
\center
\begin{tabular}{cccccc}
\hline \hline
[X/H] & $\pm \sigma_{T_{\rm eff}}$ & $\pm \sigma_{\logg}$ & $\pm \rm \sigma_{[Fe/H]}$ & $\pm \rm \sigma_{\rm v_{micro}}$ \\
\hline
O & 0.07 & 0.07 & 0.15 & 0.07 \\
Mg & 0.05 & 0.05 & 0.06 & 0.05 \\
Si & 0.00 & 0.01 & 0.00 & 0.01 \\
Ca & 0.05 & 0.05 & 0.06 & 0.04 \\
Sc & 0.09 & 0.10 & 0.09 & 0.09 \\
V & 0.07 & 0.05 & 0.06 & 0.05 \\
Cr & 0.07 & 0.05 & 0.08 & 0.06 \\
Mn & 0.05 & 0.05 & 0.05 & 0.05 \\
Co & 0.04 & 0.04 & 0.04 & 0.04 \\
Ni & 0.02 & 0.01 & 0.00 & 0.01 \\
Cu & 0.05 & 0.04 & 0.04 & 0.03 \\
Zn & 0.05 & 0.05 & 0.04 & 0.05 \\
Y & 0.06 & 0.06 & 0.09 & 0.06 \\
Zr & 0.02 & 0.03 & 0.04 & 0.03 \\
Ba & 0.08 & 0.09 & 0.07 & 0.07 \\
La & 0.09 & 0.10 & 0.11 & 0.09 \\
Ce & 0.14 & 0.14 & 0.12 & 0.14 \\
Nd & 0.10 & 0.09 & 0.09 & 0.09 \\
\hline \hline
\end{tabular}
\caption{Parameter sensitivity test using 2M10313588+3010198. We list the change in [X/H] for every element (1) as a function of changing each stellar parameter by $\pm$ the mean uncertainty in \teff~(2), \logg~(3), [Fe/H] (4), and \vmicro~(5).}
\label{tab:paramsensitivity}
\end{table}

\subsection{Uncertainties}
\label{sec:uncertainties}
There are various sources of random and systematic uncertainties that could lead to offsets between our derived abundances and those derived in APOGEE and other studies.

For example, the line selection, which we list in Table \ref{tab:lineselection}, is a source of internal scatter for the final abundance. Additionally, different line selections between the optical and the IR or between various studies also introduce systematic differences. For more details on the line selection, we refer to Appendix \ref{sec:lineselection}.

There are several other potential sources of systematics that should be noted. One such source is our solar abundance scaling which can differ by as much as 0.2 dex from the solar abundances of \citet{asplund09} such as for Co ($\Delta$ = 0.24 dex), Y ($\Delta$ = 0.25 dex), Ce ($\Delta$ = 0.22 dex), Nd ($\Delta$ = 0.20 dex), and Eu  ($\Delta$ = 0.29 dex). An additional source of systematic uncertainty is in our treatment of HFS. Taking into account the HFS for strong lines is especially important as this could lead to larger equivalent widths and overestimation of the abundance. In our analysis, this is accounted for in the creation of the model spectra from which we determine the abundance through $\chi^2$ minimization.

We also performed a parameter sensitivity test for a representative target, 2M10313588+3010198. We determined the representative target based on the mean values for the stellar parameters and their uncertainties which are $4661 \pm 95$ K for \teff, $1.63 \pm0.08$ dex for \logg, $-1.43 \pm 0.14$ dex for [Fe/H], and $1.7 \pm 0.12$ $\rm km~s^{-1}$ for \vmicro. We perturbed the stellar parameters of 2M10313588+301019 (e.g., \teff=4660 K, \logg=1.57 dex, [Fe/H]=-1.45 dex, and \vmicro=1.55 $\rm km~s^{-1}$) one by one by the average stellar parameter uncertainty, while holding all others fixed to explore how the [X/Fe] changes. The change in the abundance with respect to the change in stellar atmospheric parameter is listed in Table \ref{tab:paramsensitivity}. We note that Na and Eu were not measured for this star and are therefore not included in the table. Through this test, we find that most elements vary by $<$0.10 dex with a change in stellar parameter aside from O which is sensitive to the [Fe/H], and Ce which is sensitive to all four stellar atmospheric parameters. 

\section{Comparison with APOGEE for metal-poor stars}
\label{sec:OpticalvsIR}

\subsection{Stellar Parameters}

\begin{figure*}
\centering
\includegraphics[width=\textwidth]{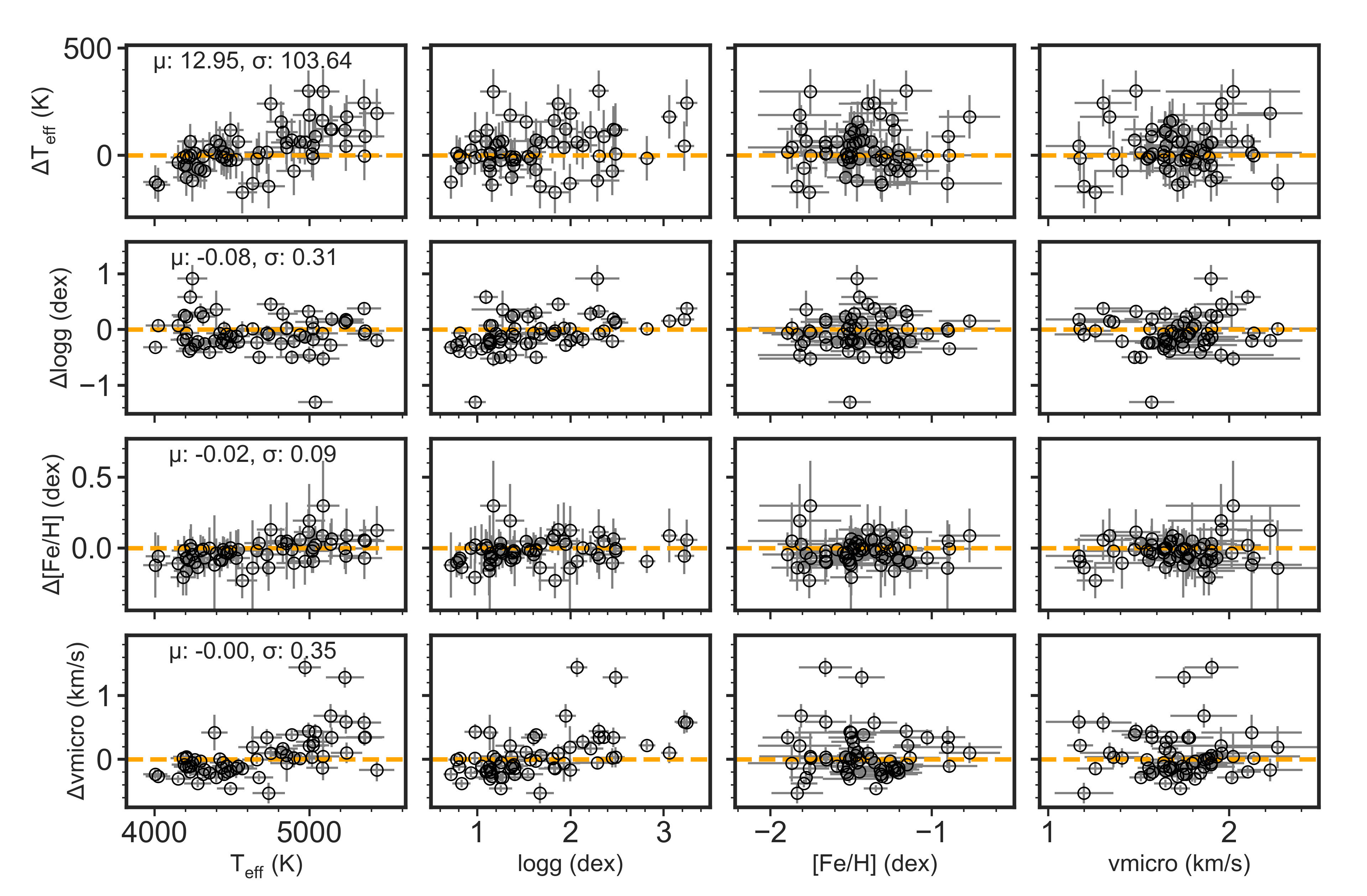}
\caption{\textit{\textbf{Comparison of stellar parameters from this study and APOGEE}}. $\Delta$ is defined as this study - APOGEE vs the values derived in this work. Starting from the left-most column going to the right, the x-axis shows \teff, \logg, [Fe/H], and \vmicro~and from the top-most row going down, the y-axis shows the $\Delta$parameter for \teff, \logg, [Fe/H], and \vmicro. We calculate the following offsets compared to APOGEE: $\Delta$\teff=12.95$\pm$103.64 K, $\Delta$\logg=-0.08$\pm$0.31 dex, $\Delta$[Fe/H]=-0.02$\pm$0.09 dex, and $\Delta$\vmicro=0$\pm$0.35 km/s.} 
\label{fig:stellarparams}
\end{figure*}
We now compare our derived stellar parameters with the parameters derived from APOGEE as illustrated in Figure \ref{fig:stellarparams}. To recall, the APOGEE stellar parameters and abundances were derived through ASPCAP \citep{garciaperez16}, which determines the best-matching full synthetic spectra that has known stellar parameters to the observed spectra. A new set of synthetic grids have been utilized for this analysis using MARCS model atmospheres \citep{gustafsson08}. We note that the reported \teff~and \logg~in APOGEE were spectroscopically-derived but were also calibrated based on the photometric \teff~from \citet{gonzalezhernandez09} for the former and asteroseismic \logg~from \citet{pins18} for the latter.

In Figure \ref{fig:stellarparams}, the $\Delta$ indicate our values minus the APOGEE values. We find a slightly higher \teff~in our analysis compared to the ASPCAP values for the same stars, with an offset of 12.95 K and scatter of 103.64 K. There is a slight positive trend between $\Delta$\teff~and \teff, as similarly seen by \citet{gonzalezhernandez09} in their comparison to the \teff~in the literature. The same work, which we based our IRFM-derived \teff~from, measure $\Delta$\teff = +50 $\pm$ 131 K for their sample of giants compared to \citet{alonso99}.

For \logg, our derived parameters are lower than the APOGEE values, with an offset of 0.08 dex and scatter 0.31 dex. There is a slight increase in $\Delta$\logg~with increasing \logg. Noticeably, there is an outlier, 2M19110434-5954152, at $\Delta$\logg$<$~1~dex. This target has an observed $M_V$ vs $V-K$ that is far from the theoretical isochrone made from the star’s [$\alpha$/Fe], [Fe/H] and an age of 10 Gyr, causing the discrepancy.

We measure [Fe/H] that are lower than APOGEE values with an offset of 0.02 dex and scatter 0.09 dex. The $\Delta$[Fe/H], shows a slight increase with \teff, akin to the trend for $\Delta$\teff~vs \teff. This follows as the measured [Fe/H] is derived from model spectra that has a prescribed \teff~from our determination.

Lastly, we compare \vmicro~and find no offset with the APOGEE values but find a scatter of 0.35 $\rm km s^{-1}$. There is a slight increasing trend in $\Delta$\vmicro~with \teff~and \logg~which, similar to [Fe/H], is because of the dependence of \vmicro~on these parameters, reflecting a similar trend. There are two outlier stars with $\Delta$\vmicro~$>$ 1 $\rm km s^{-1}$, 2M13581572+2602122 and 2M19120432-5951550, which have \vmicro~= 0.46 $\rm km s^{-1}$ and 0.47 $\rm km s^{-1}$ from APOGEE, respectively. We find that the \vmicro~derived from the optical spectra for these stars are reliable, however, because of their flat trends in \logeps~vs the reduced equivalent width.  


\subsection{Detailed Chemical Abundances from the Optical and IR}
\label{sec:chem_OvsIR}

\begin{figure*}
\centering
\includegraphics[width=\textwidth]{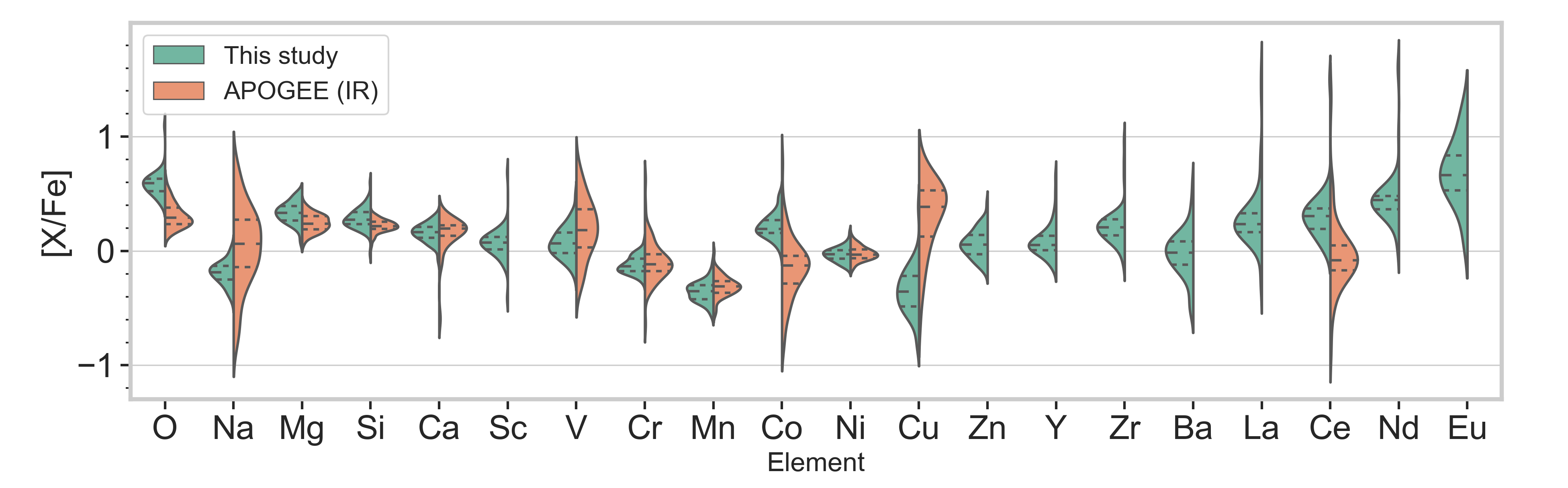}
\caption{\textit{\textbf{Violin plot summarizing the [X/Fe] abundances}}. We show the abundance distributions of 21 elements measured in this study (left, gold) side-by-side with the abundances from APOGEE where available (right, orange),ordered by increasing atomic number, to highlight the differences in the offsets and scatters in our sample from the optical and from the IR. We tabulate the abundance means and the scatters derived in this study in Table \ref{tab:stats}. We note that the optical and IR samples for a given element do not necessarily have the same number of stars.} 
\label{fig:scatter}
\end{figure*}

\begin{table*}
\begin{center}
\begin{tabular}{c|ccc|ccc|cc}
\hline \hline
(1) & (2) & (3) & (4) & (5) & (6) & (7) & (8) & (9) \\
$\rm [X/Fe]$ & $\rm \mu_{optical}$ (dex) & $\rm \sigma_{optical}$  (dex) & $\rm N_{optical}$ & $\rm \mu_{IR}$  (dex) & $\rm \sigma_{IR}$  (dex) & $\rm N_{IR}$ & $\rm median_{[X/Fe], O-IR}$  (dex) & $\rm \sigma_{[X/Fe], O-IR}$  (dex) \\ 
\hline
Mg & 0.33 & 0.09 & 60 & 0.24 & 0.08 & 62 & 0.09 & 0.09 \\
O & 0.58 & 0.11 & 45 & 0.31 & 0.11 & 56 & 0.30 & 0.15 \\
Si & 0.29 & 0.09 & 62 & 0.22 & 0.05 & 62 & 0.07 & 0.08 \\
Ca & 0.16 & 0.08 & 62 & 0.15 & 0.17 & 62 & 0.01 & 0.18\\
\hline
Mn & -0.36 & 0.09 & 59 & -0.31 & 0.09 & 61 & -0.04 & 0.11 \\
Cr & -0.12 & 0.08 & 62 & -0.09 & 0.18 & 62 & -0.03 & 0.20 \\
Co & 0.22 & 0.11 & 57 & -0.15 & 0.27 & 60 & 0.36 & 0.25 \\
Ni & -0.03 & 0.06 & 62 & -0.02 & 0.06 & 62 & -0.01 & 0.08 \\
Zn & 0.06 & 0.12 & 62 & - & - & - & - & - \\
\hline
Na & -0.19 & 0.10 & 49 & 0.04 & 0.35 & 56 & -0.24 & 0.38 \\
Sc & 0.07 & 0.13 & 62 & - & - & - & - & - \\
V & 0.07 & 0.14 & 60 & 0.19 & 0.26 & 47 & -0.11 & 0.27 \\
Cu & -0.35 & 0.21 & 47 & 0.31 & 0.31 & 59 & -0.60 & 0.40 \\
\hline
Ba & -0.02  & 0.21 & 58 & - & - & - & - & - \\
La & 0.29  & 0.28 & 59 & - & - & - & - & - \\
Ce & 0.31 & 0.22 & 47 & -0.04 & 0.32 & 47 & 0.35 &  0.25\\
Nd & 0.46  & 0.25 & 62 & - & - & - & - & - \\
Y & 0.08 & 0.14 & 60 & - & - & - & - & - \\
Zr & 0.22 & 0.17 & 56 & - & - & - & - & - \\
Eu & 0.69 & 0.28 & 14 & - & - & - & - & - \\
\hline \hline
\end{tabular}
\caption{Summary statistics for the [X/Fe] trends as shown by the violin plot in Figure \ref{fig:scatter}.  We tabulate the mean [X/Fe], scatter in [X/Fe], and the number of stars where the element, X, is measured for the optical data (columns 2-4) and APOGEE (columns 5-7). We also report the median offset and the scatter in $\rm [X/Fe]_{optical} - [X/Fe]_{IR}$ (columns 8 and 9). }
\label{tab:stats}
\end{center}
\end{table*}

In this work, we have selected accreted halo stars based solely on their chemistry using the IR-derived abundances for [Mg/Mn] and [Al/Fe] from APOGEE DR16. We then derived elemental abundances in follow-up optical spectra for species spanning different families i.e. $\alpha$, Fe-peak, light and odd-Z, and more notably neutron-capture (both $s$ and $r$). 

From this analysis, we are able to compare the optical vs IR abundances for 13 elements we have in common with APOGEE DR16, namely O, Mg, Si, Ca, V, Cr, Mn, Co, Ni, Cu, and Ce, as shown in the violin plot in Figure \ref{fig:scatter} and listed in Table \ref{tab:stats}. Of these elements, O, Co, Na, Cu, and Ce show offsets $>$0.15 dex. For some elements such as Co, Cu, and Ce, this is partially due to the abundances we derived from twilight which were different by 0.20 dex from the solar abundance reported from \citet{asplund09}. Cr, Co, Na, V, Cu, and Ce show optical-IR scatters $>$0.20 dex, which are mostly driven by the scatter in the ASPCAP abundances as seen in Figure \ref{fig:scatter}. \citet{jonsson18} similarly compared the APOGEE DR13 and DR14 abundances with independent analyses in the optical and measure a median offset and scatter of 0.05 dex and 0.15 dex, respectively. On the other hand, our median offset and scatter are -0.01 dex and 0.18 dex, respectively; we note that our sample is significantly metal-poorer, and we include Cu (offset by -0.60$\pm$0.40 dex) and Ce (offset by 0.35$\pm$ 0.25 dex) which are not in the \citet{jonsson18} analysis, yet our median offset and scatter are comparable to theirs. 

Some elements are better measured in the optical while some are better measured in the IR. For example we are able to derive the neutron-capture element abundances for seven species, in comparison to only one (Ce) in the IR data from APOGEE for our sample of accreted halo stars. On the other hand, we could only, if even possible, poorly derive abundances for N and Al with our optical spectra and therefore did not include those results, whilst APOGEE has good measurements for them. Extensive discussions about individual elements can be further found in the recent review of \cite[section 4.3]{jofre19}. 
Al is one of the elements in APOGEE that our target selection was based on, therefore confirming and quantifying the systematics between its optical and IR abundances is critical. However, we were unable to determine accurate Al abundances from the optical spectra. First, the atomic information of the Al lines 6696.0~\AA\ and 6698.7~\AA\ have been flagged as possibly unreliable \citep{jofre19, heiter2021atomic}. Second, 3D and non-LTE effects are different for each line, in particular in metal-poor stars. \cite{nordlander17} studied these effects for the 2 lines discussed here and at least one included in APOGEE (16763~\AA), illustrating this point.  Third, in most of the cases of this particular study, they fall on the echelle order gaps. For the 11 stars in which the line was detected, the lines either fall between orders making the normalisation in that region very uncertain or the spectra were too noisy and abundances unreliable. 

Additionally, determining Al in metal-poor stars is not straightforward indeed. The recent work of \cite{roederer21} shows how for metal-poor stars the two optical lines used here are rarely detected, and if detected, yield abundances that are significantly higher than from blue or near UV lines, making Al abundances difficult to interpret. Hence, to properly compare IR with optical Al abundances for metal-poor stars remain subject of future studies. 


\subsubsection{A comprehensive chemical dataset for accreted halo stars}

Understanding the chemical abundances of accreted halo stars from both the optical and the IR is therefore largely complementary and beneficial and gives access to as many elements as possible. Because of this, in the following analysis, we show mostly the optically-derived abundances for our sample but for some elements we consider instead the APOGEE ASPCAP values, keeping in mind their systematic differences as outlined in this section. Specifically, we used the IR values for the following elements (1) O because the derived abundance in the optical is enhanced ($\mu$ = 0.58 dex) which is contrary to what is expected for stars with likely accreted origins (enhanced [O/Fe] were similarly found by \citealt{jonsson18} in their independent optical spectra analysis of APOGEE stars) and (2) Si because its total scatter in the optical is larger by 0.04 dex compared to the scatter in the IR, unlike the other elements measured in the optical. This could possibly be due to our line selection in Si that introduces an internal scatter of 0.08 dex, compared to that in the IR (e.g., 0.05 dex). 


\section{Chemical Abundance Patterns of blob stars}
\label{sec:abundances}
Here, we discuss the detailed chemical abundances of the blob stars alongside different stellar populations.  
We show the [X/Fe] vs [Fe/H] trends for these potential accreted stars as open stars colored black if from the optical and gray if from the IR. We group the elements into their different families: $\alpha$ elements (Figures \ref{fig:global_alpha} and \ref{fig:alpha}), Fe-peak (Figure \ref{fig:ironpeak}), light and odd-Z (Figure \ref{fig:lightodd}), and neutron-capture (Figure \ref{fig:ncapture}). Where available, we include data for the Milky Way thin disc, thick disc, and halo \citep{bensby14,battistini15,battistini16,reddy03,reddy06,yong13} as purple diamonds, low-$\alpha$ accreted halo stars \citep{Nissen10,nissen11,fishlock17} as gold pentagons, and LMC \citep{pompeia08,vdswealmen13} as green triangles. We also mark [X/Fe] = 0 dex and [Fe/H] = 0 dex (i.e. solar metallicity) with grey dashed lines to guide the eye. We note that these comparison data do not have overlap with our sample, and therefore caution the reader that there are likely systematic offsets that exist between our study and those in the literature as discussed in Sections \ref{sec:uncertainties} and \ref{sec:chem_OvsIR}. We also refer back to these sections regarding a discussion of the offsets between our optical abundances and the APOGEE IR abundances in interpreting our results.

\subsection{$\alpha$ elements: Mg, O, Si, Ca}

\begin{figure}
\centering
\includegraphics[width=0.48\textwidth]{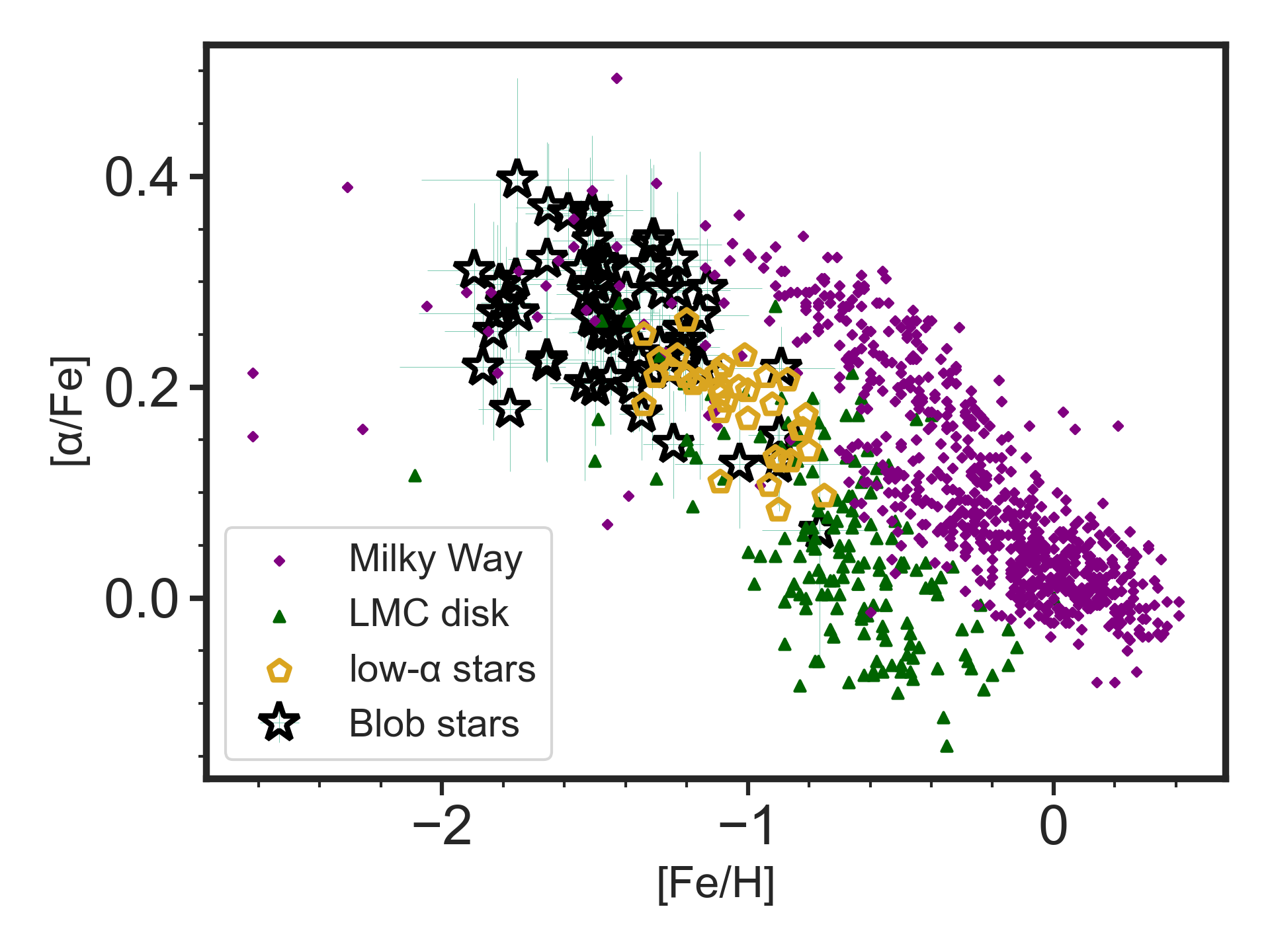}
\caption{\textbf{\textit{Total [$\alpha$/Fe] trend}}. We calculated [$\alpha$/Fe] from the abundances of Mg, Si, and Ca for the blob targets in the optical (black stars) and compare to the Milky Way disc (purple diamonds, \citealt{bensby14}), low-$\alpha$ stars (teal pentagons, \citealt{Nissen10}), and the LMC (green triangles, \citealt{vdswealmen13}). The blob stars show lower [$\alpha$/Fe] abundance compared to the Milky Way at [Fe/H] $>$-1.5 dex and similar trends to the LMC and low-$\alpha$ stars.} 
\label{fig:global_alpha}
\end{figure}

\begin{figure*}
\centering
\includegraphics[width=\textwidth]{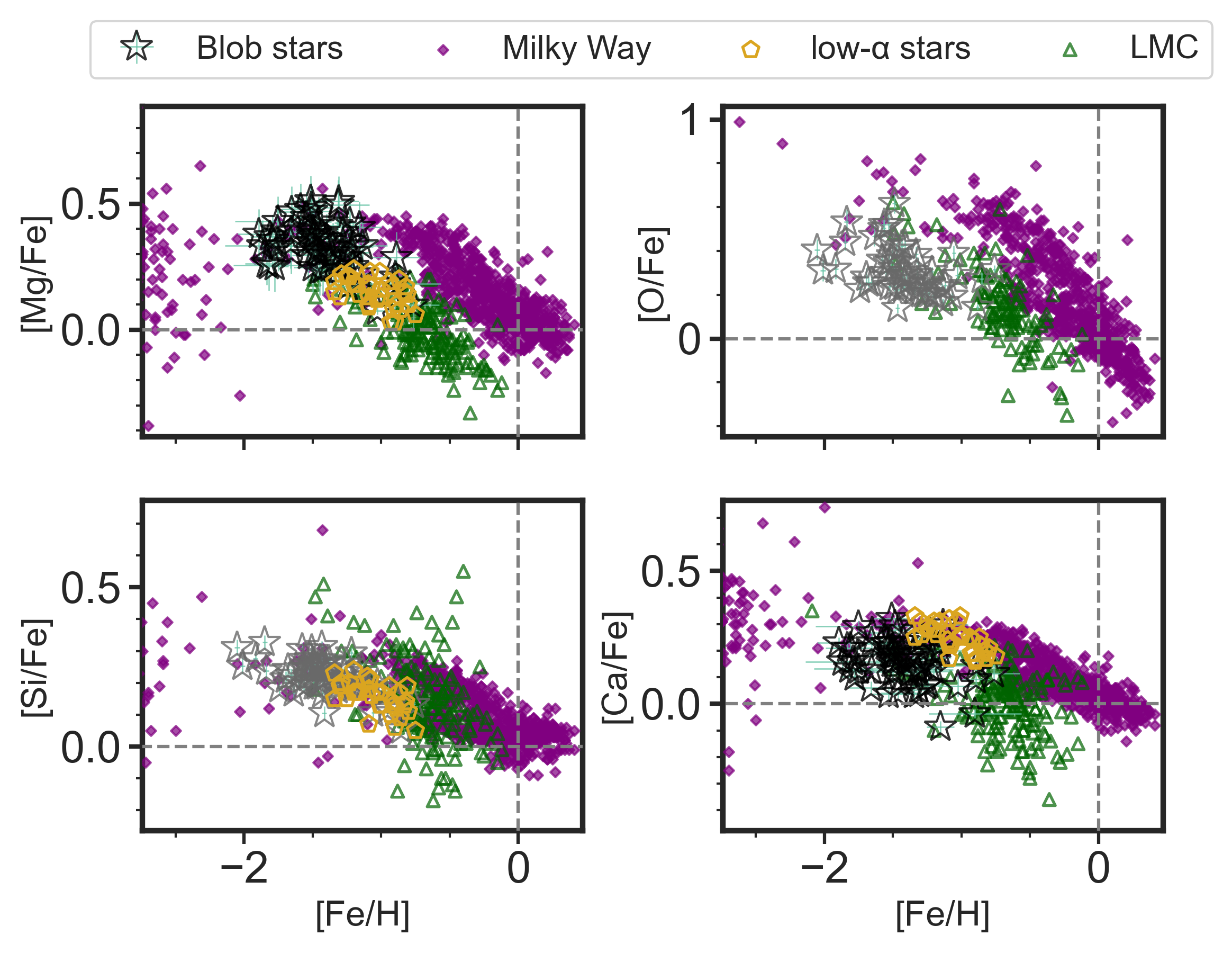}
\caption{\textbf{\textit{$\alpha$ elements}}. The [X/Fe] abundance ratios in the $\alpha$ family of elements---Mg, O, Si, and Ca---against [Fe/H] for the blob stars in the optical (black stars) and IR (gray stars) as well as other comparison data. These samples are shown as follows: purple diamonds for the Milky Way \citep{yong13,bensby14}, gold pentagons for accreted stars \citep{Nissen10}, and green triangles for LMC \citep{vdswealmen13}. 
We mark [X/Fe] = 0 and [Fe/H] = 0 with dashed grey lines to guide the eye. } 
\label{fig:alpha}
\end{figure*}

We show the [$\alpha$/Fe] vs [Fe/H] for our sample in Figure \ref{fig:global_alpha}, where $\alpha$ is the averaged abundance for Mg, Ca, and Si derived from the high-resolution optical spectra. We measure lower [$\alpha$/Fe] for the blob stars compared to the Milky Way disk stars at the same [Fe/H], up to [Fe/H]$\sim$-1.5 dex. 
The global [$\alpha$/Fe] trend show similarities with the LMC and the low-$\alpha$ stars from \citet{Nissen10} where their [Fe/H] overlap. We next discuss the individual [X/Fe] for the $\alpha$ elements.  
\linebreak
\linebreak
\textbf{Mg:} The [Mg/Fe] trend of blob stars is similar to what \citet{nissen11} found for their low-$\alpha$ sample, where the low-$\alpha$ population is distinct from the high-$\alpha$ population up to [Fe/H]=-1.5 dex. The blob stars trend is similar to that of the LMC as well, where the former appears to be an extension of the latter's trend to lower [Fe/H]. At -1.4 dex $<$ [Fe/H] $<$ -0.8 dex, where stars from our sample overlap with the comparison stars from LMC, we see overlapping range in [Mg/Fe]. For the higher [Fe/H] end of our sample (i.e. [Fe/H] $>$ -1.5 dex), the [Mg/Fe] values are systematically lower compared to the Milky Way disc (e.g. \citealt{bensby14}), in line with other previous studies of the accreted halo population (e.g. \citealt{Nissen10,hawkins15,helmi18,hayes18,mackereth19}).
\linebreak
\linebreak
\textbf{O:} The [O/Fe] trend in the IR is lower than the Milky Way disc trend from \citet{bensby14}, similar to what is found for our Mg results as well as for the previous studies of low-$\alpha$ stars \citep{hawkins15,hayes18}. This is reasonable given that both O and Mg are hydrostatic $\alpha$ elements which are created during hydrostatic burning that occurs in shells within the cores of massive stars. Where there is overlap in [Fe/H], the blob stars and the LMC stars from \citet{vdswealmen13} similarly overlap in their [O/Fe] abundance, though the blob stars' trend appears to have  different and shallower slope. 
\linebreak
\linebreak
\textbf{Si:} The [Si/Fe] trend in the IR is similar to the low-$\alpha$ stars trend from \citet{Nissen10} and to the LMC at the same [Fe/H]. Previous studies have noted that the low-$\alpha$ and high-$\alpha$ stars are less separated in [Si/Fe] (e.g. \citealt{hawkins15,hayes18}) compared to other $\alpha$ elements in APOGEE, which we similarly find in our sample comparing to Milky Way disc stars from \citet{bensby14}.  
\linebreak
\linebreak
\textbf{Ca:} The [Ca/Fe] trend for the blob stars show the most similarity with the LMC trend of all the $\alpha$ elements. It is lower than both the trends for the Milky Way (disc and halo) and for the low-$\alpha$ stars from \citet{Nissen10}. Like Si, Ca also does not separate the in-situ Milky Way stars from the accreted halo stars as well as Mg and O do \citep{Nissen10,hayes18}, especially at the [Fe/H] regime in this study. However, at the higher [Fe/H] end of our sample, i.e. [Fe/H] $>$ -1.0 dex, the blob stars have lower [Ca/Fe] values compared to the thick disc stars from \citet{bensby14}. 

\subsection{Fe-peak elements}
\begin{figure*}
\centering
\includegraphics[width=\textwidth]{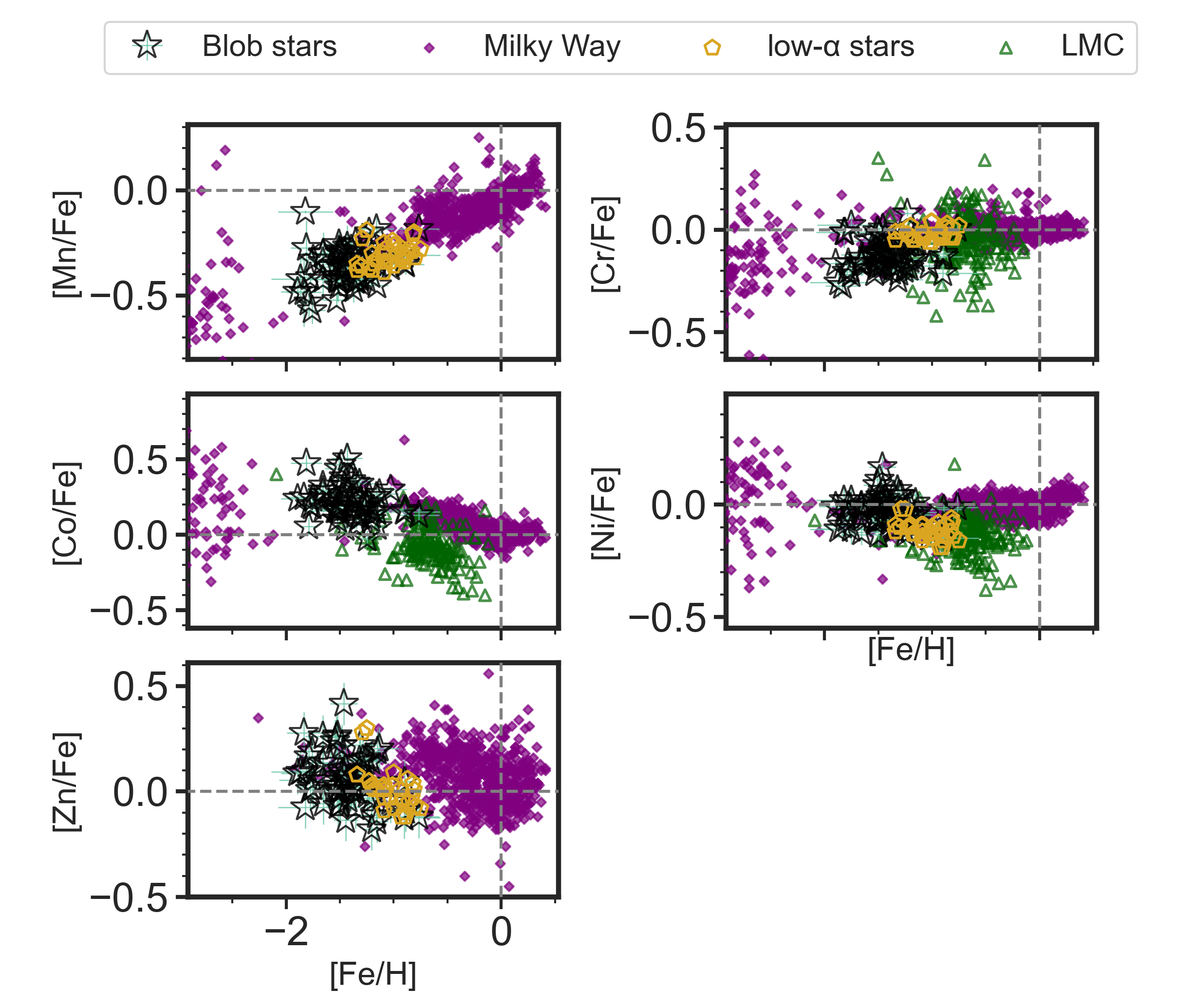}
\caption{\textbf{\textit{Fe-peak elements}}. The [X/Fe] abundance ratios for Fe-peak elements (Mn, Cr, Co, Ni, and Zn) against [Fe/H] for the blob stars in the optical (black stars) as well as other comparison data shown as follows: purple diamonds for the Milky Way \citep{yong13,bensby14,battistini15}, gold pentagons for accreted stars \citep{nissen11}, and green triangles for LMC \citep{vdswealmen13}. 
We mark [X/Fe] = 0 and [Fe/H] = 0 with dashed grey lines to guide the eye.} 
\label{fig:ironpeak}
\end{figure*}

Fe-peak elements (Mn, Cr, Co, Ni, Zn), in contrast to $\alpha$ elements are predominantly produced and dispersed via SNIa \citep{iwamoto99}. 
These elements generally track the Fe abundance although other production mechanisms like Si-burning in core-collapse supernovae (SNe) cause deviations from this general trend  \citep{kobayashi06}. 
\linebreak
\linebreak
\textbf{Mn:} The [Mn/Fe] we measure for the blob stars in optical decreases with decreasing [Fe/H]. The sample of low-$\alpha$ stars from \citet{nissen11} also agree with this [Mn/Fe] trend at the higher [Fe/H] end. \citet{nissen11} and \citet{hayes18} find that their sample of low-$\alpha$ (Mg) stars are indistinguishable in the [Mn/Fe] vs [Fe/H] plane from in-situ Milky Way stars, which is shown as well with our comparison to Milky Way disc and halo trends.  The comparison studies (e.g. \citet{nissen11} for the low-$\alpha$ stars and \citet{battistini15} for the Milky Way disc stars) both take into account HFS for Mn from \citet{prochaska00}.
\linebreak
\linebreak
\textbf{Cr:} The [Cr/Fe] trend has a slight decrease with lower [Fe/H] and is systematically lower from the solar value (i.e. [Cr/Fe] = 0 dex) by 0.12 dex. This trend is slightly lower compared to that of the Milky Way and LMC, and would be even lower by 0.10 dex if we used the solar Cr abundance from \citet{asplund09} rather than the one derived from our twilight spectra. At the higher [Fe/H] end of the sample, the Cr abundances overlap with the low-$\alpha$ stars from \citet{nissen11}, where they find that the low- and high-$\alpha$ stellar populations are not distinct from each other. 
\linebreak
\linebreak
\textbf{Co:} The [Co/Fe] trend for the blob stars decreases with higher [Fe/H]. The [Co/Fe] trend for the blob stars and the LMC seem to follow the same trend where (1) they have similar [Co/Fe] values where their [Fe/H] overlap and (2) they have similar slopes with [Fe/H] that are shallower than the Milky Way's. \citet{vdswealmen13} account for Co HFS for the LMC using data from \citet{fuhr88} while \citet{battistini15} use \citet{prochaska00} for the Milky Way disc stars and both works, similar to our analysis, determine the abundance from the best-fit model spectra, especially when the element is affected by HFS. 
\linebreak
\linebreak
\textbf{Ni:} The [Ni/Fe] trend for the blob stars decreases with increasing [Fe/H]. \citet{hayes18} noted that of the Fe-peak elements in their study, [Ni/Fe] separates the low-Mg (accreted halo) and high-Mg (in-situ) populations the most. At [Fe/H] $\geq$ -1.5 dex, the [Ni/Fe] trend for the blob stars is below the Milky Way disc trend at a given [Fe/H], similar to previous studies \citep{Nissen10,hawkins15,hayes18,mackereth19}. 
Interestingly, the [Ni/Fe] abundances overlap for our sample of stars and for the LMC at the same [Fe/H].
\linebreak
\linebreak
\textbf{Zn:} The [Zn/Fe] trend for the blob stars is enhanced at [Fe/H] $<$-1.5 dex and decreases with increasing [Fe/H], reaching solar values at [Fe/H] = -1 dex. The blob stars share a similar [Zn/Fe] trend with the low-$\alpha$ stars from \citet{Nissen10}, where the two samples overlap at [Fe/H] $>$-1.5 dex. \citet{Nissen10} also noted that the [Zn/Fe] trend for their low-$\alpha$ stars, i.e. decreasing Zn with increasing [Fe/H], is different from the trend they find for the high-$\alpha$ (in-situ) stars, i.e. constant Zn with [Fe/H]. 
For comparison, the Milky Way disc stars have solar [Zn/Fe] abundance for the most part, changing to a slightly above-solar but flat trend at [Fe/H] $<$ -1.0 dex, distinct from what we find for our sample.

\subsection{Light and Odd-Z elements}
\begin{figure*}
\centering
\includegraphics[width=\textwidth]{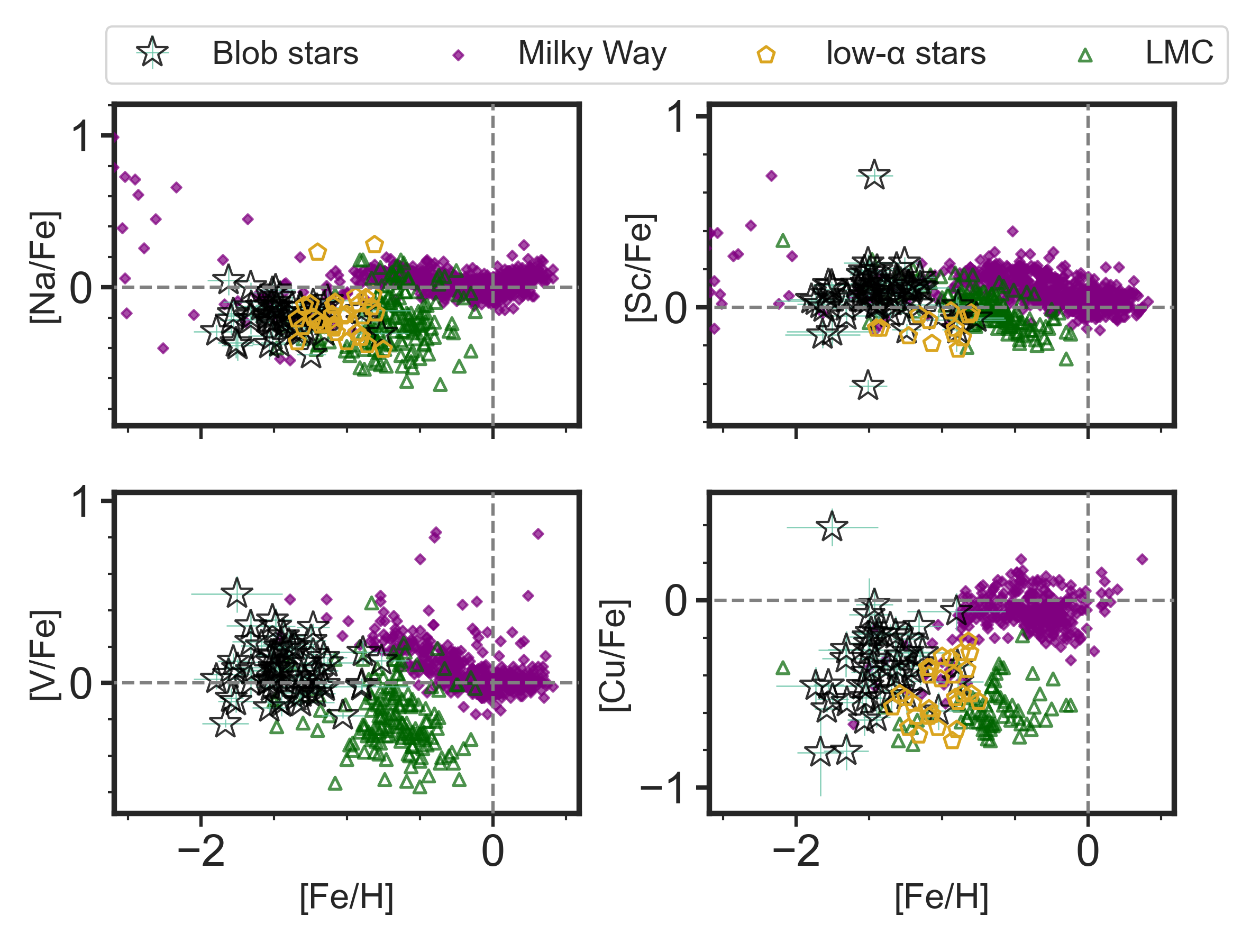}
\caption{\textbf{\textit{Light and odd-Z elements}}. The [X/Fe] abundance ratios for light and odd-Z elements (Na, Sc, V, and Cu) against [Fe/H] for the blob stars in the optical (black stars) as well as other comparison data shown as follows: purple diamonds for the Milky Way \citep{yong13,bensby14,battistini15, reddy03,reddy06}, gold pentagons for accreted stars \citep{Nissen10,nissen11,fishlock17}, and green triangles for LMC \citep{vdswealmen13}. 
We mark [X/Fe] = 0 and [Fe/H] = 0 with dashed grey lines to guide the eye.} 
\label{fig:lightodd}
\end{figure*}

The production of Odd-Z elements depend on the neutron excess from $\rm ^{22}Ne$ and are therefore, for the most part, metallicity-dependent except for certain elements (e.g. V) \citep{kobayashi20}.
Na is mainly produced in core-collapse SNe and is sensitive to the initial C in the star forming gas \citep{kobayashi06}. At higher [Fe/H], AGB stars are also expected to produce some Na \citep{nomoto13}. Sc is mainly produced in massive stars during C and Ne burning and is eventually dispersed through core-collapse SNe \citep{woosley02}. It is therefore seen to follow other $\alpha$ elements in its trend with metallicity. 
\linebreak
\linebreak
\textbf{Na:} The [Na/Fe] for the blob stars in the optical shows a flat trend with [Fe/H] with a mean abundance of -0.19 dex and error of 0.05 dex. In comparison, the same set of stars in APOGEE have typical uncertainties of $\sim$0.20 dex, highlighting why deriving the abundance in the optical for certain elements, such as Na, is necessary. 
\citet{Nissen10} find a depleted and flat [Na/Fe] trend for their low-$\alpha$ stars compared to their high-$\alpha$ stars as well, similar to what we find for our blob stars compared to Milky Way stars from \citet{bensby14} at [Fe/H]~$\gtrsim$~-1.2~dex. At lower [Fe/H] the Milky Way disc stars overlap with our sample.  The blob stars [Na/Fe] abundances are also very similar to the LMC's where there is overlap in [Fe/H]. 
\linebreak
\linebreak
\textbf{Sc:} The [Sc/Fe] trend for the blob stars is essentially flat with [Fe/H]. In comparison, the LMC stars from \citet{vdswealmen13} show a flat trend that turns over and decreases at higher [Fe/H]. \citet{fishlock17} specifically followed up on the accreted stars from \citet{Nissen10} and derived [Sc/Fe] abundances to compare to an earlier study (i.e.  \citealt{nissen00}). \citet{nissen00} found that the [Sc/Fe] of the low-$\alpha$ stars is lower compared to that of the high-$\alpha$ stars at the same [Fe/H] for the range -1.4 $<$ [Fe/H] $<$ -0.7 dex. At the higher [Fe/H] end of our sample, the blob stars overlap with the stars from \citet{fishlock17} and both are lower than the Milky Way disk trend at the same [Fe/H]. However, at the same [Fe/H], some of our program stars also have higher [Sc/Fe] abundances, and in general the mean [Sc/Fe] for the blob stars is 0.07 dex. 
\linebreak
\linebreak
\textbf{V:} We find a flat [V/Fe] trend for the blob stars with a mean [V/Fe] = 0.07 dex. Where there is overlap in metallicity i.e. [Fe/H]$>$-1.5 dex, the [V/Fe] abundance trend for our sample also greatly resembles the LMC trend from \citet{vdswealmen13}, both distinct from the Milky Way disc \citep{battistini15}. Both comparison studies also take into account HFS, namely from \citet{martin88} for the LMC and from \citet{prochaska00} for the Milky Way disc stars. Taking into account the HFS effect is important for V. For example, as \citet{vdswealmen13} illustrated for the LMC, they measure [V/Fe] that is higher by $\approx$0.22 dex without the HFS corrections.
\linebreak
\linebreak
\textbf{Cu:}  We find depleted [Cu/Fe] abundances for the blob stars with mean of -0.35 dex, where the trend is essentially flat for [Fe/H]$>$-1.5 dex but is decreasing with decreasing [Fe/H] below this value. Our Cu result agrees with that of \citet{nissen11} as well as the more recent study from \citet{matsuno20}, where both show [Cu/Fe] abundances below 0 dex for low-$\alpha$ accreted halo stars. Taking into account the difference between our solar Cu abundance and that of \citet{asplund09} would put the blob stars' [Cu/Fe] at an even lower trend. 
\citet{vdswealmen13} similarly find [Cu/Fe] $<$ 0 dex for the LMC. 
The blob stars, the low-$\alpha$ stars, and the LMC all show lower Cu abundances than the Milky Way \citep{reddy03,reddy06}. All comparison studies also take into account the HFS for Cu, i.e., \citet{prochaska00} for the low-$\alpha$ stars, \citet{bielski75} for the LMC, and Kurucz models for the Milky Way stars from \citet{reddy03,reddy06}.

\subsection{Neutron-capture elements}
\begin{figure*}
\centering
\includegraphics[width=\textwidth]{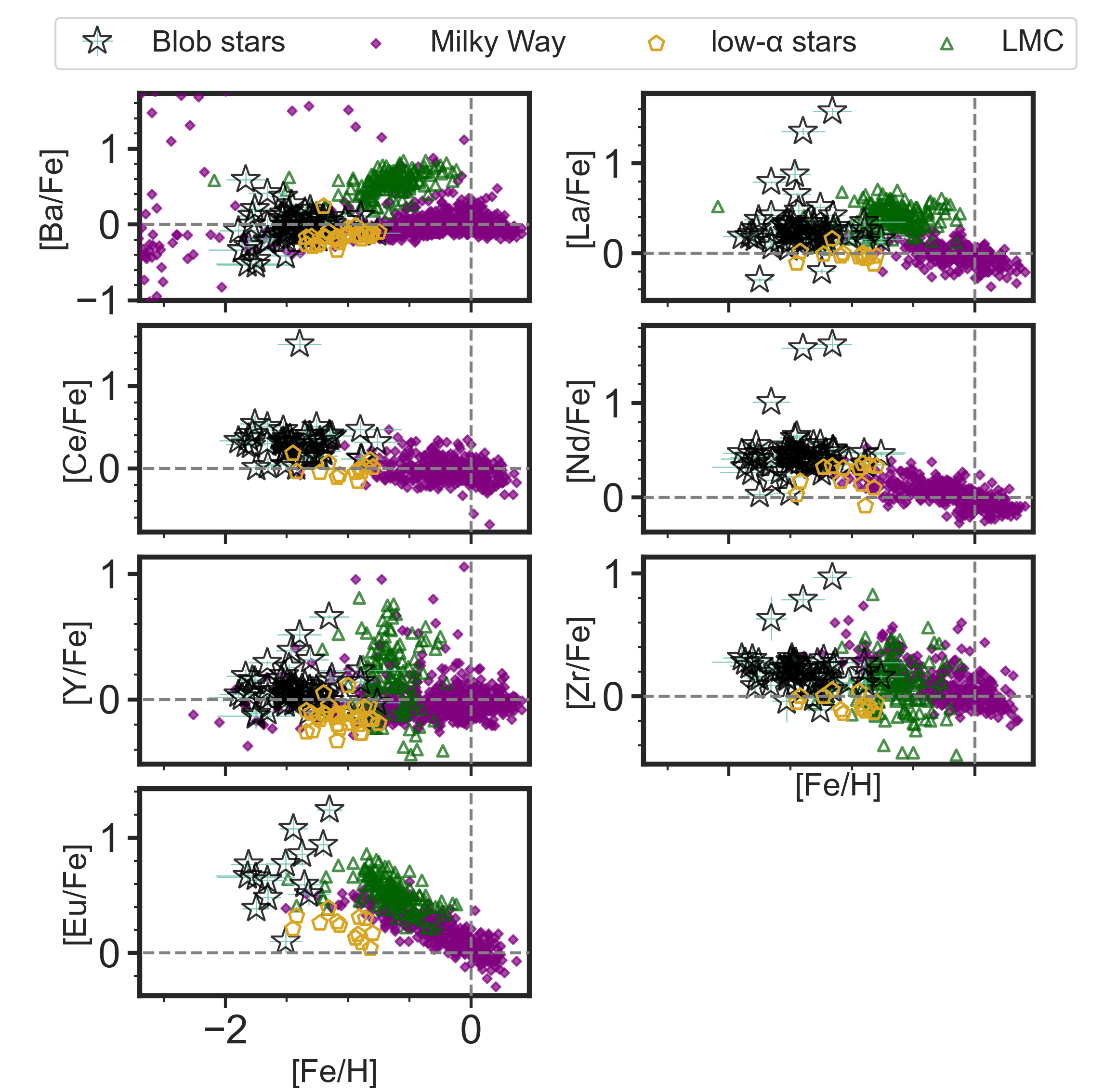}
\caption{\textbf{\textit{Neutron-capture elements}}. The [X/Fe] abundance ratios for neutron-capture elements---Ba, La, Ce, Nd (heavy-\textit{s} process), Y, Zr (light-\textit{s} process), and Eu (r-process)---against [Fe/H] for the the blob stars in the optical (black stars) and the IR (gray stars) as well as other comparison data shown as follows: purple diamonds for the Milky Way \citep{yong13,bensby14,battistini16}, gold pentagons for accreted stars \citep{nissen11,fishlock17}, and green triangles for LMC \citep{vdswealmen13}. 
We mark [X/Fe] = 0 and [Fe/H] = 0 with dashed grey lines to guide the eye.} 
\label{fig:ncapture}
\end{figure*}

Neutron-capture elements are heavier than Fe and made through a slow ($s$) or rapid ($r$) capture of neutrons. The main difference is that the $s$-process neutron capture happens slower than the $\beta$-decay while the $r$-process captures many neutrons at timescales faster than the $\beta$-decay. The envelopes of AGB stars are considered the main source of $s$-process \citep{busso99}, but a weaker component also comes from massive stars related to $^{22}$Ne neutron production \citep{pignatari10,frisch16}. On the other hand, the production site of $r$-process elements are less known, though core-collapse SNe and neutron star mergers provide viable pathways (e.g. \citealt{kasen17}). Ba, La, Ce, and Nd are heavy $s$-process elements while  Y, and Zr are light $s$-process elements, with heavy and light pertaining to the atomic number \citep{busso01}. Eu on the other hand is considered a pure $r$-process elements, contributing 97\% of the Eu present in the sun \citep{sneden08}. 
\linebreak
\linebreak
\textbf{Ba:}The blob stars have a flat [Ba/Fe] trend with [Fe/H] centered around solar values (mean of -0.02 dex) and a scatter that increases with lower [Fe/H] irrespective of the SNR. On the other hand, the low-$\alpha$ stars from \citet{nissen11} have [Ba/Fe] below solar where they find that the high and low-$\alpha$ halo stars overlap in [Ba/Fe]. The sample of accreted stars from \citet{matsuno21} has an elevated range in [Ba/Fe] i.e., 0~$<$~[Ba/Fe]~$<$~1~dex compared to our sample with -0.5~$<$~[Ba/Fe]~$<$~0.6~dex at the same [Fe/H] range. 
\citet{aguado20} also measure [Ba/Fe] for GES stars and find Ba abundances that span a large range of above-solar values from GALAH DR3 \citep{buder20}, though for the stars which they derived the abundances themselves, they measure solar [Ba/Fe], similar to what we find. Compared to our sample, the LMC has elevated [Ba/Fe] centered on $\sim$~0.5 dex with [Ba/Fe] increasing with [Fe/H]. The MW disc similarly has [Ba/Fe] centered at 0 dex. All the comparison studies account for HFS as well, namely \citet{mcwilliam98} for the low-$\alpha$ stars and \citet{rutten78} for the LMC.
\linebreak
\linebreak
\textbf{La:} The blob stars have a flat [La/Fe] trend with [Fe/H], similar to the low-$\alpha$ stars from \citet{fishlock17}. However, our stars are offset higher by 0.3~dex compared to their sample centered at [La/Fe] = 0~dex, where they find no difference in the [La/Fe] abundance of high and low-$\alpha$ stars. \citet{matsuno20} similarly find elevated [La/Fe] abundances for their sample of accreted stars centered at $\sim$0.4 dex. \citet{matsuno21} further studied GES stars using GALAH DR3 data and find elevated [La/Fe] abundances as well. They likewise find that La does not separate accreted vs in-situ populations. The blob stars' [La/Fe] trend exhibit similarities with the LMC, as the LMC also shows a flat and elevated [La/Fe] trend with [Fe/H] compared to the Milky Way disc. In contrast, the Milky Way disc has lower [La/Fe] trend that decreases with [Fe/H]. 
\linebreak
\linebreak
\textbf{Ce:} The [Ce/Fe] trend for the blob stars is flat and centered at [Ce/Fe] = 0.31 dex. On the other hand, the low-$\alpha$ accreted halo stars from \citet{fishlock17} have subsolar [Ce/Fe] that is flat with [Fe/H].The same authors note that there is no difference in the Ce abundances of their sample of high and low-$\alpha$ stars. \citet{matsuno20} also find elevated Ce values for their low-$\alpha$ stars sample, with the [Ce/Fe] trend centered at $\sim$0.2 dex. 
\linebreak
\linebreak
\textbf{Nd:} The blob stars show a flat [Nd/Fe] trend with [Fe/H], similar to the low-$\alpha$ stars from \citet{fishlock17} but shifted to higher values. Our sample is centered on 0.46 dex while theirs is centered on 0.20 dex. However, the [Nd/Fe] trend for the blob stars would be shifted lower by 0.20 dex and closer to the trend from \citet{fishlock17} if we were to use solar abundance from \citet{asplund09}. These authors find no difference in the [Nd/Fe] of their sample of high and low-$\alpha$ stars, though the Nd for the low-$\alpha$ stars have a higher scatter. \citet{matsuno20} also measure Nd for their sample of accreted stars and find similarly flat and elevated values at [Nd/Fe] = 0.40 dex. They likewise find no difference in the [Nd/Fe] trend with metallicity for their sample of accreted versus in-situ stars. The Milky Way disc stars from \citet{battistini16}, on the other hand show a lower, and decreasing [Nd/Fe] trend with higher [Fe/H]. This comparison Milky Way data had HFS accounted for from \citet{roederer08}.
\linebreak
\linebreak
\textbf{Y:} The [Y/Fe] trend for the blob stars is flat with [Fe/H] and elevated with a mean of 0.08 dex.  For comparison, \citet{nissen11}, \citet{aguado20}, and \citet{matsuno20} measure below solar, below solar, and solar values for [Y/Fe], respectively, with the low-$\alpha$ stars from  \citet{nissen11} also showing a flat [Y/Fe] vs [Fe/H] trend. Adopting Y solar abundance from \citet{asplund09} however would decrease the mean abundance by $\sim$0.25~dex and put our sample at the same [Y/Fe] range as these studies. The Milky Way stars \citep{bensby14} show a flat [Y/Fe] trend with [Fe/H] that similarly lie $<$0 dex. The LMC, on the other hand, has an elevated [Y/Fe] vs [Fe/H] trend compared to the blob stars.
\linebreak
\linebreak
\textbf{Zr:} The blob stars exhibit a flat [Zr/Fe] vs [Fe/H] trend centered at 0.22 dex. This trend is offset higher compared to the low-$\alpha$ accreted stars sample from \citet{fishlock17} which has [Zr/Fe]~$\sim$~0~dex. The LMC [Zr/Fe] trend from \citet{vdswealmen13} is similarly flat and elevated. On the contrary, the Milky Way disc from \citet{battistini16} shows decreasing [Zr/Fe] with higher [Fe/H].
\linebreak
\linebreak
\textbf{Eu:} Though we have a sample of 62 accreted halo stars, we are only able to measure Eu for a handful of them. We find enhancement in Eu for the blob stars that is flat and centered at [Eu/Fe]=0.69 dex. Using Eu solar abundance from \citet{asplund09} would shift the trend down by 0.3 dex. The sample of low-$\alpha$ stars from \citet{fishlock17} are offset lower but still with super-solar values centered at 0.30 dex which slightly decreases with increasing [Fe/H]. These authors find that the low-$\alpha$ stars have higher [Eu/Fe] compared to the high-$\alpha$ stars that have in-situ origins. \citet{matsuno20}, \citet{matsuno21}, and \citet{aguado20} all find enhanced Eu abundances for their respective samples of low-$\alpha$ accreted stars, pointing to the prevalence of r-process in their progenitor.  It is worth noting that r-process sites are rare and so the r-process abundance trends in low-mass systems are mostly due to stochastic effects (e.g., \citealt{ji16}). \citet{matsuno21} additionally find that [Eu/Fe] separate in-situ versus accreted material very well, and [Eu/Mg] even more so. \citet{vdswealmen13} measure Eu abundances for LMC stars that decrease with increasing [Fe/H], similar to what \citet{fishlock17} find for their sample, but unlike the trend that we see for the blob stars that seem to increase with [Fe/H]. However, it is harder to comment more definitively on the [Eu/Fe] trend because we do not have many stars with measured Eu abundance and for those that do, they show an overall high scatter $\sim$0.28 dex. Both the accreted stars and the LMC exhibit higher [Eu/Fe] abundances compared to in-situ Milky Way stars from \citet{battistini16}. These comparison studies all similarly take into account HFS i.e., from \citet{lawler01} for the accreted stars from \citet{aguado20} and \citet{fishlock17}, and the Milky Way stars from \cite{battistini16}.


\section{Discussion}
\label{sec:discussion}

\subsection{Neutron-capture element abundance ratios}
\label{sec:ncapture_ratios}

\begin{figure*}
\centering
\includegraphics[width=\textwidth]{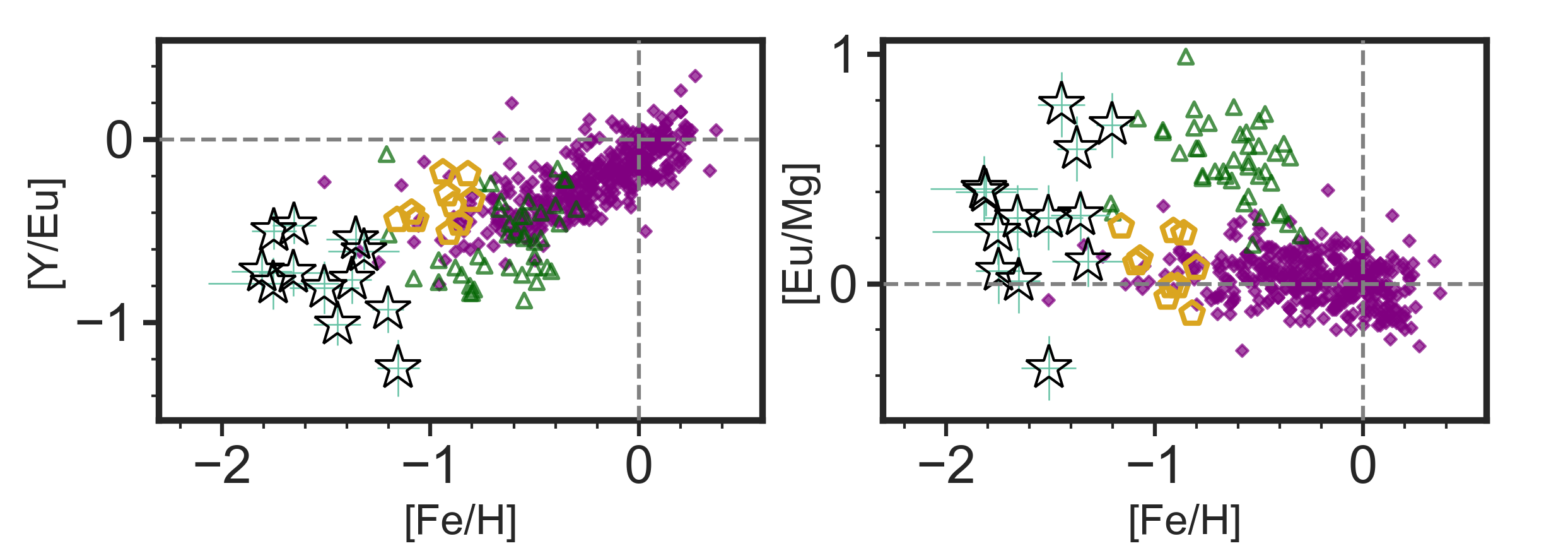}
\caption{\textbf{\textit{Element ratios with Eu}}. [Y/Eu] and [Eu/Mg] trends with [Fe/H] for the blob sample (black stars) as well as other comparison data shown as follows: purple diamonds for the Milky Way \citep{bensby14,battistini16}, gold pentagons for accreted stars \citep{nissen11,fishlock17}, and green triangles for LMC \citep{vdswealmen13}. 
We mark [X/Fe] = 0 with dashed grey lines to guide the eye.}
\label{fig:eu_ratios}
\end{figure*}

\begin{figure*}
\centering
\includegraphics[width=\textwidth]{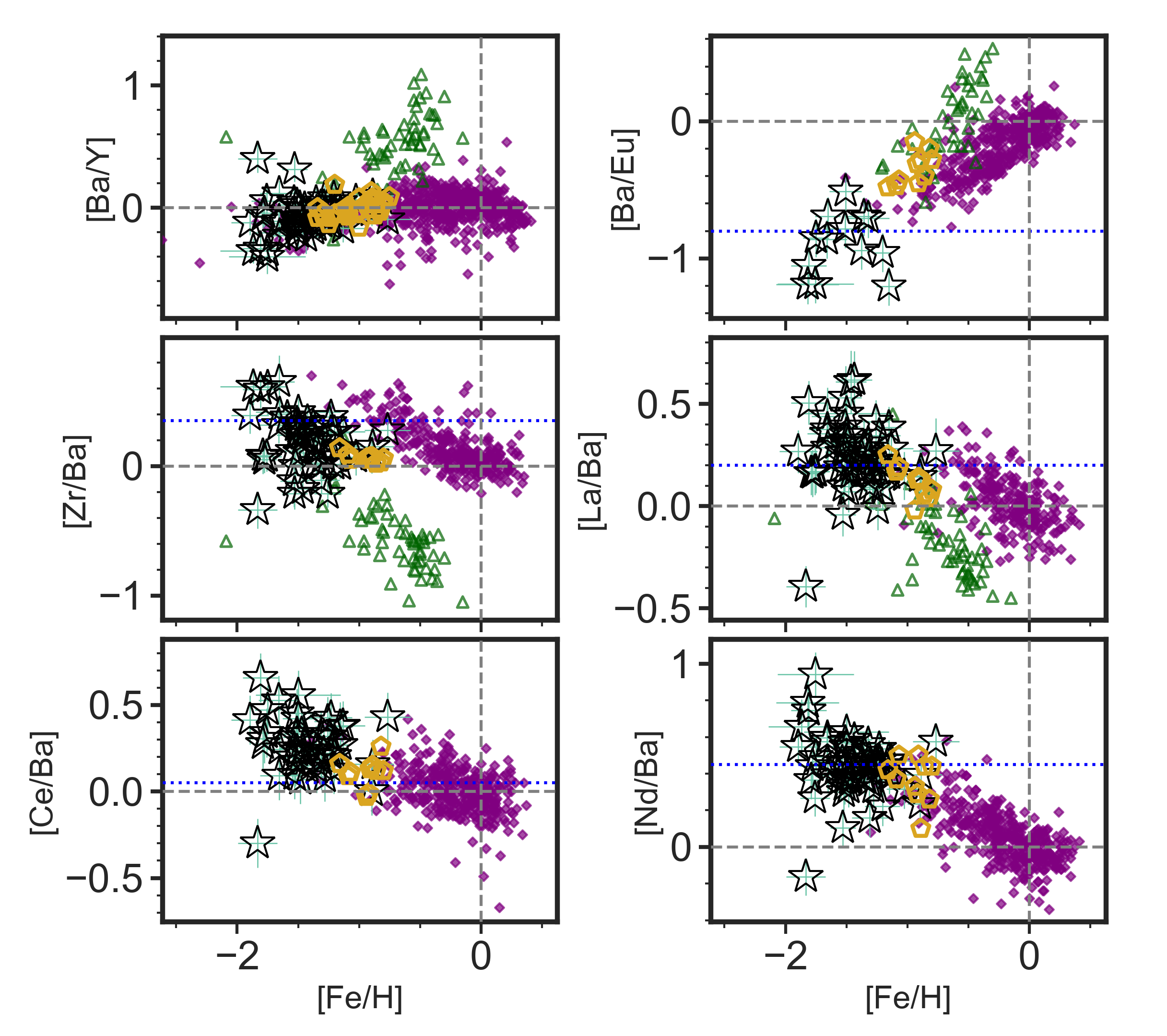}
\caption{\textbf{\textit{Element ratios with Ba}}. Element to element ratio with Ba as a function of [Fe/H] for the blob sample (black stars) as well as other comparison data shown as follows: purple diamonds for the Milky Way \citep{bensby14,battistini16}, gold pentagons for accreted stars \citep{nissen11,fishlock17}, and green triangles for LMC \citep{vdswealmen13}. 
We mark [X/Fe] = 0 and [Fe/H] = 0 with dashed grey lines to guide the eye. We also show a pure r-process ratio with a dotted blue line taken from \citet{bisterzo14}.}
\label{fig:ba_ratios}
\end{figure*}

\begin{figure}
\centering
\includegraphics[width=0.45\textwidth]{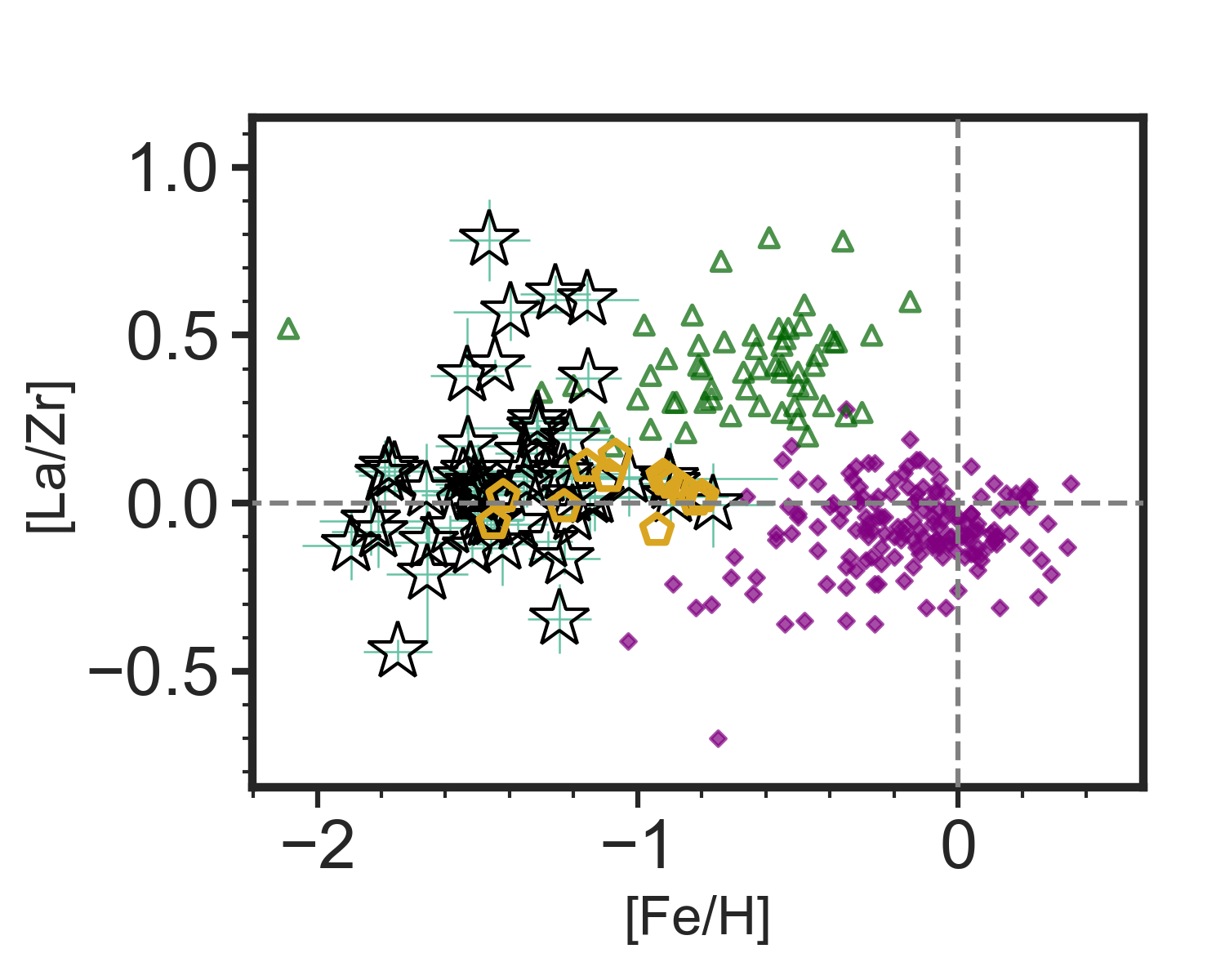}
\caption{[La/Zr] vs [Fe/H] the blob sample (black stars) as well as other comparison data shown as follows: purple diamonds for the Milky Way \citep{bensby14,battistini16}, gold pentagons for accreted stars \citep{fishlock17}, and green triangles for LMC \citep{vdswealmen13}. 
We mark [X/Fe] = 0 with dashed grey lines to guide the eye.}
\label{fig:lazr}
\end{figure}

We take a deeper look into the neutron-capture abundance ratios with respect to each other to understand better which processes dominate the chemical evolution of our sample of accreted stars. Specifically, we highlight [Y/Eu] and [Eu/Mg] as a function of [Fe/H] in Figure \ref{fig:eu_ratios}, abundance ratios with respect to Ba in Figure \ref{fig:ba_ratios}, and [La/Zr] vs [Fe/H] in Figure \ref{fig:lazr}. The figure legends follow those of Figures \ref{fig:global_alpha}, \ref{fig:alpha}, \ref{fig:ironpeak}, \ref{fig:lightodd}, and \ref{fig:ncapture}. We combined data sets with overlapping sample selection and methods to compare our results to other stellar populations, i.e., \citet{bensby14} and \citet{battistini16} for the Milky Way, and \citet{Nissen10}, \citet{nissen11}, and \citet{fishlock17} for low-$\alpha$ stars. We also include stars from the LMC from \citet{vdswealmen13}.

These combinations of elements have been previously explored by other studies on accreted stars \citep[e.g.][]{nissen11,fishlock17,aguado20,matsuno20,matsuno21} and compared to trends within dwarf galaxies \citep[e.g.][]{venn04,tolstoy09} because they are an indicator of the relative contributions of s- and r-processes in the creation of these elements. This is especially important to discuss for dwarf galaxies because (1) r-process nucleosynthesis is very rare, and so its stochasticity in these lower-mass systems can greatly affect r-process enrichment, and (2) a delayed s-process enrichment suggests lower star formation rate.

The GCE modeling from \citet{bisterzo14} find the following solar s-process contributions: 72\% for Y, 66\% for Zr, 85\% for Ba, 76\% for La, 83\% for Ce, 58\% for Nd, and 6\% for Eu. In Figure \ref{fig:ba_ratios}, we show where the stars lie (dotted blue line) should they have been purely made through r-process, which is adapted from \citet{battistini16} using the GCE model from \citet{bisterzo14}, where they had a sample of Milky Way stars within the metallicity range -1.4 $<$ [Fe/H] $<$ 0.4 dex. We note that our comparison is done with solar s-process predictions though our sample has lower [Fe/H]. Nonetheless, we show the resulting pure r-process prediction from this as a reference point as previous studies have done.  
 
\subsubsection{[Y/Eu] and [Eu/Mg]}
\label{sec:eu_ratios}

The blob stars show depleted [Y/Eu] at -0.76$\pm$0.22 dex, which is lower compared to the LMC, Milky Way and low-$\alpha$ stars. However, we note that our sample is more metal-poor and we barely have overlapping [Fe/H] with these other stellar populations. Since the scatter is large, it is hard to tell if there is a trend in [Y/Eu] with [Fe/H], although as s-process should start contributing at later times, the [Y/Eu] trend is expected to increase with [Fe/H] as is seen for the Milky Way. \citet{fishlock17} and \citet{matsuno20} find that at the same [Fe/H], the in-situ Milky Way stars have higher [Y/Eu], suggesting the delayed nature of s-process production in the accreted population. This may be similar to what we see for the blob stars with [Fe/H]$>$-1.5~dex compared to the Milky Way stars from \citet{bensby14} and \citet{battistini16}, but this could possibly be due to systematic differences. Present day dwarf spheroidals have also been observed to have a lower [Y/Eu] compare to the Milky Way halo \citep{venn04,tolstoy09}. Y is a product of the weak s-process that depends on $\rm ^{22}Ne$ for neutrons, which in turn depends on the CNO abundance. Similar to \citet{matsuno20} and \citet{nissen11}, we find that the weak s-process is inefficient in the progenitor of the blob stars because of their depleted [Y/Eu], suggesting a lower star formation rate. 

The [Eu/Mg] trend for the blob stars is elevated with a mean abundance of 0.29$\pm$0.29 dex, slightly higher than low-$\alpha$ stars from \citet{Nissen10} and \citet{fishlock17} and the Milky Way \citep{bensby14,battistini16}, but lower compared to the LMC \citep{vdswealmen13}. \citet{matsuno20,matsuno21} similarly find elevated [Eu/Mg] at $\sim$0.30 dex for their GES stars compared to in-situ stars ([Eu/Mg] = 0 dex) at the same [Fe/H]. These authors posited that the observed [Eu/Mg] for GES is either due to (1) the delayed production of Eu through r-process with neutron star mergers or (2) underproduction of Mg through a top-light initial mass function (IMF). Their chemical evolution modeling supports the former scenario, in agreement with our results, which we discuss in detail in the following section comparing element abundances to Ba. Indeed, a more recent study by \citet{naidu21} proposed that the r-process enrichment from NSM is delayed by 0.5 to 1 Gyr to explain their GES Eu abundance. 

\subsubsection{Element abundance ratios with Barium}
\label{sec:ba_ratios}

In this section, we compare the element abundance ratios with respect Ba. We find an increasing [Ba/Y] with [Fe/H] for the blob stars, similar to what \citet{nissen11} find for their sample of low-$\alpha$ stars. Our sample has a a mean [Ba/Y] = -0.06 $\pm$ 0.14 while \citet{fishlock17} report [Ba/Y] = 0.01 $\pm$ 0.06 dex for the low-$\alpha$ population. The chemical evolution modeling from \citet{nissen11} posit that the increasing trend in [Ba/Y] for the low-$\alpha$ stars supports a progenitor that had a lower rate of chemical evolution, mostly enriched by SNIa with a delayed s-process enrichment from low-metallicity low-mass AGB stars. Such high [Ba/Y] ratio is also seen in dwarf spheroidals around the Milky Way and is attributed to lower rate of chemical evolution \citep{tolstoy09}.


Following from the [Ba/Y], \citet{fishlock17} explored the [La/Zr] trend for low-$\alpha$ stars as Ba and La are both heavy s-process elements produced through the main s-process in the envelopes of AGB stars, while Y and Zr are both light s-process elements produced through the weak s-process in massive stars. We similarly show in Figure \ref{fig:lazr} the [La/Zr] as a function of [Fe/H] for the blob stars. We find a mean of 0.07 dex for our sample and though there is large scatter, the majority of the blob stars follow the same trend as the low-$\alpha$ stars from \citet{fishlock17}. The same authors used predictions from \citet{fishlock14} to explore the [La/Zr] for AGB stars of different masses and find that at [Fe/H] = -1.2 dex, the super-solar [La/Zr] ratios are more consistent with enrichment by low mass (i.e., 1-3 $\rm M_{\odot}$) AGB stars. 

[Ba/Eu] is a good diagnostic in chemical evolution as Ba is mainly produced through the s-process while Eu is considered a pure r-process elements, with only 6\% of Eu being produced in s-process at solar and low [Fe/H] (e.g., -0.3 dex) \citep{bisterzo11,bisterzo14}. The [Ba/Eu] trend for the blob stars support a pure r-process enrichment, with a mean of -0.89$\pm$0.21, in agreement with the theoretical abundance ratio for r-process \citep{bisterzo14,battistini16}. On the other hand, the low-$\alpha$ stars from \citet{fishlock17} are offset higher at -0.39$\pm$0.12 dex, though they also have higher [Fe/H]. They also find that the in-situ stars have higher [Ba/Eu], further supporting the r-process enhancement scenario for the low-$\alpha$ stars compared to the Milky Way.  In comparison, the LMC has a higher [Ba/Eu] compared to the Milky Way at the same [Fe/H], alluding to the larger contribution of s-process. This heavy s-process enhancement is similarly seen in other present day dwarf galaxies \citep{letarte10,hill19} which further shows that the neutron-capture element enrichment in the blob progenitor and present day Milky Way dwarfs are dominated by different processes. The GES stars from \citet{aguado20} similarly exhibit depleted [Ba/Eu] at -0.7 dex and find that the trend is flat for [Fe/H]$<$-1.5 dex, and increases at higher metallicity, signaling the start of s-process enrichment. To confirm this trend seen by the authors, we require Eu abundance measurements from higher metallicity blob stars.

The rest of the element abundance ratios with Ba in Figure \ref{fig:ba_ratios} i.e., [Zr/Ba] at 0.19$\pm$0.21 dex, [La/Ba] at 0.24$\pm$0.16 dex, [Ce/Ba] at 0.27$\pm$0.16 dex, and [Nd/Ba] at 0.44$\pm$0.16 dex further support that the blob stars are r-process enhanced, which is different compared to present day dwarfs like the LMC that have greater s-process contribution because of an extended star formation compared to GES.

\subsection{Comparison to previous studies of low-$\alpha$ stars}
\label{sec:comp_studies}

We further compare the abundance patterns of the blob stars to those of low-$\alpha$ accreted halo stars from \citet{Nissen10,nissen11}, \citet{fishlock17}, and \citet{matsuno20}. This comparison comes with caveats in that the data, sample of stars, and analyses are different among these studies and ours. 
All three comparative studies did a line-by-line differential analysis with a low-metallicity standard star and used solar abundances from \citet{asplund05,asplund09} while we derived our abundances with respect to the Sun and used solar abundances derived from our own twilight spectra. However, even with these differences, we can still compare the relative trends of the accreted halo stars from each study. We note that there are absolute differences as well, but caution that this could be due to a number of effects (e.g., line selection, continuum normalization, NLTE, HFS treatment etc.).  

Stellar populations in the halo that have different [$\alpha$/Fe] abundances and kinematics have been observed in earlier studies (e.g. \citealt{nissen97,fulbright02,gratton03}). However, whether this abundance distribution is continuous or bimodal was not known until \citet{Nissen10,nissen11} and \citet{schuster12} derived high precision abundances and  orbital integration for their sample of inner halo stars. We therefore focus our comparison with \citet{Nissen10} and \citet{nissen11} that derived abundances for $\alpha$, Fe-peak, light/odd-Z, and neutron-capture elements. They selected low metallicity dwarf stars in the solar neighborhood that have halo kinematics (i.e. $V_{\rm total}>$ 180 $\rm km$  $s^{-1}$). In addition to the two populations separating in [$\alpha$/Fe]-[Fe/H], they also differ in their Ni and Na trends with the low-$\alpha$ population having lower values compared to the high-$\alpha$ population. These trends are similar to our result (Figure \ref{fig:lightodd} and \ref{fig:ironpeak}), indicating that our blob stars are chemically similar to the low-$\alpha$, accreted stars from \citet{Nissen10,nissen11}. They also found low values and high scatter for Cu in their low-$\alpha$ population, distinct from the high-$\alpha$ trend which is offset higher with lower scatter. We observe the same behavior as their low-$\alpha$ stars for our sample's Cu abundances. \citet{kobayashi06, kobayashi20} showed that Cu production depends on the excess neutrons from $\rm ^{22}Ne$ which is produced from $\rm ^{14}N$ during the CNO cycle; \citet{nissen11} posit that the trend they found, similar to ours, is due to these stars forming from gas enriched by SNIa and Cu produced by CNO-poor massive stars. This picture is also supported by the blob stars' lower abundances in Na and Ni compared to the Milky Way at the same [Fe/H], as these elements are similarly made in massive stars. 

\citet{fishlock17} followed up on the sample from \citet{Nissen10} and derived abundances for neutron-capture elements (e.g. Zr, La, Ce, Nd, and Eu) as well as Sc for 27 Galactic dwarf stars. They investigated the neutron-capture element abundances of these stars to constrain the mass range of AGB stars that contributed to the chemical enrichment of their progenitor. 
Among their neutron-capture elements, they found that light-$s$ process element Zr separates the low-$\alpha$ stars from high-$\alpha$ in-situ stars the most. 
The measured Zr abundance for our sample is elevated compared to their study, though the relative trends are similar where both show an essentially flat trend with [Fe/H].  On the other hand, the heavy-$s$ process elements i.e. La, Ce, and Nd do not have distinct trends between these two populations. Similar to \citet{Nissen10}, their chemical evolution modeling of these n-capture abundances point to a time-delay of low-mass AGB stars' contribution. Though we find offsets between our derived [X/Fe] for the neutron-capture elements and those from this study, the neutron-capture element-to-element ratios actually agree between these two samples, as discussed in Section \ref{sec:ncapture_ratios} and shown in Figure \ref{fig:lazr} pointing to similar chemical evolution scenarios. 

Lastly, we compare to the recent study by \citet{matsuno20} that, similar to this work, observed low-$\alpha$ accreted halo stars in the red giant branch. Specifically, their sample consisted of 7 metal-poor stars, 7 $\alpha$-rich stars, 9 $\alpha$-poor stars, and 3 very $\alpha$-poor stars. They measured abundances for 23 different elements, similarly spanning groups that are of interest to our study, and in addition, derived masses for these stars with asteroseismology. They found subsolar values for Y as well as Cu, similar to \citet{nissen11}, a result of the time delay in the light-s process element production. In contrast, we measure slightly above solar values for our Y abundances, but subsolar values for the Cu abundances, similar to \citet{matsuno20}. However, we do confirm an underproduction of Y (Figure \ref{fig:eu_ratios}) and delay in s-process enrichment (Figure \ref{fig:ba_ratios}) as discussed in the previous section when looking at neutron-capture abundance ratios.    

Very recently, there have been works by \citet{matsuno21}, \citet{aguado20}, and \citet{naidu21} that focus on the r-process in GES stars where all found elevated Eu, highlighting the r-process enhancement in the GES system. There are various sites for r-process production (e.g., electron-capture supernovae or magnetorotationally-driven supernovae, \citet{cescutti14}) but the simple chemical evolution model from \citet{matsuno21} best agree with a neutron star merger scenario (e.g. \citealt{rosswog18}) for their sample of GES stars. The 14 stars where we measure Eu from similarly show Eu enhancement and the neutron-capture element abundance ratio discussion in Section \ref{sec:ncapture_ratios} support the dominance of the r-process in the progenitor of the blob stars. 

From this comparison, we put forth that the detailed abundance patterns for our sample of blob stars generally agree with these previous studies of low-$\alpha$ stars and 
suggest that they were accreted from a dwarf galaxy with a lower rate of chemical evolution compared to the Milky Way. It is quite noteworthy that the detailed chemical abundances of our sample of likely accreted halo stars, which were selected purely through chemical tagging, have good overall agreement with these studies of low-$\alpha$, accreted stars that were selected differently through kinematics, or sometimes in combination with metallicity.

\subsection{Comparison to Milky Way dwarf galaxies}
\label{sec:comp_dwarfgals}

The previous section, as well as more recent studies suggest that these low-$\alpha$ blob stars were accreted from a massive dwarf galaxy with stellar mass in the range of $\rm 8.5 < log M_{\odot} < 9.0$ \citep{mackereth19, fernandezalvar18} and a total mass of $\rm \sim 10^{11} M_{\odot}$ \citep{das20} during the epoch of disc formation, 8-10 Gyr ago. \citet{das20} find a range of ages (8 to 13 Gyr) of stars belonging to the blob, and argue that it is hard to disentangle a single vs multiple progenitor scenario, especially as these stars were accreted earlier in the history of our Galaxy. However, as pointed out by the same authors, not very many massive systems can exist around the Milky Way (see \citealt{rodriguezpuebla13}). In this section, we compare the detailed chemical abundances of blob stars to those of massive Milky Way dwarf galaxies. 

First we compare to the LMC, Milky Way's most massive satellite with a dynamical mass of $\sim \rm 10^{11} M_{\odot}$ \citep{erkal20}. This is a relevant comparison as mass estimates for the GES system roughly compare to that of LMC. From Section \ref{sec:abundances} and using LMC abundances from \citet{vdswealmen13}, we showed that the $\alpha$ (Mg, Si, Ca), Fe-peak (Cr, Ni, Co), light/odd-Z element (Na, V, Cu), and neutron-capture (La, Y, Zr) element trends for the blob stars bear great similarity with the trends from LMC.  
However, a closer look at the neutron-capture element ratios show that the blob stars have a larger r-process enrichment compared to the LMC. These comparisons indicate that although the $\alpha$, Fe-peak, and light/odd-Z production may be similar for the blob progenitor and the LMC, their heavy-element production are distinct from each other. However, this could well may be a [Fe/H] effect; it is interesting to speculate how the blob track would compare to the LMC if it survived to the present day.

Next, we compare to the Milky Way dwarf galaxy Sagittarius \citep{ibata94}, which is currently being disrupted by the Galaxy's potential. From its stars' velocity dispersion, \citet{law10} estimated an initial bound mass of $\rm 6.4 \times 10^{8}~M_{\odot}$ for Sagittarius before it was stripped off stars by the Milky Way. It is interesting to think of these comparison satellite galaxies as follows: the LMC as an intact massive satellite, Sagittarius as a currently disrupted massive satellite, and GES as a fully disrupted massive satellite. \citet{mcwilliam13} similarly found low [$\alpha$/Fe], [Na/Fe], [Al/Fe], 
and [Cu/Fe] for the Sagittarius dwarf stars compared to the Milky Way halo. However, they posit that this is due instead to a top-light IMF instead of a SNIa time-delay. 

Fornax is another massive Milky Way dwarf satellite with mass $\rm 1.6\times10^{8}~M_{\odot}$ \citep{lokas09}. \citet{letarte10} investigated the detailed chemical abundance of Fornax stars and found, similar to this study, $\alpha$ element abundance trends that are lower than the Milky Way's at a given [Fe/H]. They also measured low Ni and Na, akin to what we see for our blob stars. It is important to make a distinction however, that the stars in Fornax from \citet{letarte10} are young (2 to 6 Gyr old) compared to the blob stars in our sample. 

\citet{tolstoy09} and \citet{nissen11} discussed in length the [$\alpha$/Fe] vs [Fe/H] of different Milky Way dwarf galaxies and where their respective [Fe/H]$\rm _{knee}$ and [Fe/H]$\rm _{max}$ occur. The former is an indication of star formation rate i.e. when SNIa started enriching the system's gas while the latter is related to how long star formation proceeded in the galaxy. 
Lower mass dwarf galaxies have lower [Fe/H]$\rm _{knee}$ and [Fe/H]$\rm _{max}$ (e.g. \citealt{koch08}). Meanwhile, the dwarf galaxies we have compared to in this section have higher values for either parameter, and sometimes both. Sagittarius has [Fe/H]$\rm _{knee}$ = -1.3 dex and [Fe/H]$\rm _{max}$ = 0 dex \citep{sbordone07,carretta10}, and Fornax has [Fe/H]$\rm _{max}$ = -0.6 dex \citep{letarte10}. The LMC has an interestingly low [Fe/H]$\rm _{knee}$ at $\leq$-2.2 dex \citep{nidever20}. On the other hand, \citet{nissen11} did not find a [Fe/H]$\rm _{knee}$ for their low-$\alpha$ because of a lack of metal poor stars with [Fe/H] $<$ -1.5 dex, though they found a [Fe/H]$\rm _{max}$ $\approx$ -0.75 dex . In contrast, we cover a larger sample of metal-poor stars but a sample extending to even lower metallicities i.e. [Fe/H] $<$ -2.0 is needed in order to fully characterize where the plateau is.

To summarize, these comparisons to Milky Way dwarfs suggest that the blob stars were likely accreted from a massive dwarf galaxy that had a lower rate of chemical evolution that the Milky Way.

\section{Summary}
\label{sec:summary}
In this work, we combine optical and IR data to understand the detailed chemical abundance patterns of accreted halo stars. We have obtained detailed chemical abundances for 62 accreted halo stars using high-resolution (R$\sim$40,000) optical spectra which were chemically selected from APOGEE DR16 using [Mg/Mn] vs [Al/Fe], a combination of chemical abundances that separates the in-situ Milky Way stars from accreted stars that form a "blob" in this plane. We obtained abundances for 20 elements spanning the $\alpha$, Fe-peak, light, odd-Z, and neutron-capture groups of elements. 
This work outlines the detailed chemical abundances of these accreted halo stars in both the optical and the IR to understand these individual elements' abundance trends with respect to the Milky Way, its satellites, and previously-known low-$\alpha$ stars. Investigating neutron-capture element abundances, which are harder to access in the IR, is especially important for systems with lower rate of chemical evolution i.e. dwarf galaxies less massive than the Milky Way. 


We find that for the most part, the optical and IR abundances agree but some elements like O, Co, Na, Cu, and Ce have trends that highlight the differences in deriving abundances in the optical and the IR. We confirm that the $\alpha$ element trends for the blob stars, both individual and total, are lower than the Milky Way disk at the same metallicity for [Fe/H] > -1.5 dex. This, together with our sample's depleted abundances in Ni, Na, and Cu suggest that their progenitor system had a lower star formation rate or lower rate of chemical evolution. 
This is similar to what previous studies of low-$\alpha$ stars \citep{Nissen10,nissen11,fishlock17,matsuno20} found, which is even more interesting given that we had a completely different sample selection. We measure abundances for seven neutron-capture elements (e.g. Ba, La, Ce, Nd, Y, Zr, Eu) and find that they are similarly enhanced compared to the low-$\alpha$ stars sample from \citet{matsuno20}, \citet{matsuno21}, and \citet{aguado20}, and that they exhibit relative trends that agree with \citet{Nissen10,nissen11} and \citet{fishlock14}. Additionally, we explored the neutron-capture abundance ratios and similarly find that our blob sample is r-process enhanced and had delayed and inefficient s-process enrichment. When coupled with the [$\alpha$/Fe] ratios, this supports a star formation scenario slower than the Milky Way's, that was interrupted before significant contribution by AGB stars to the chemical enrichment. This overall agreement with previous studies of accreted halo stars proves that our independent selection is noteworthy. Our work has shown that chemically selecting accreted stars in [Mg/Mn] vs [Al/Fe] is a powerful tool that could be used for upcoming surveys that peer deeper into the halo, especially if the kinematics are uncertain.  
Lastly, comparing to Milky Way satellites (e.g.,\citealt{tolstoy09,letarte10,vdswealmen13,mcwilliam13}), we find that the blob stars exhibit detailed chemical abundances in the $\alpha$, Fe-peak, and light/odd-Z elements in line with those of massive Milky Way dwarf galaxies, especially the LMC, but heavy-element abundance patterns that point to contributions from different production mechanisms.  

This work presents the detailed chemical abundances for accreted halo stars
and opens up many avenues for follow-up investigation. Galactic chemical evolution modeling would provide a better way of understanding the contribution and timescales of processes and stars that enriched the progenitor of these accreted stars. As discussed in Section \ref{sec:comp_dwarfgals}, it would be worthwhile to find the knee in the [$\alpha$/Fe] vs [Fe/H] plane for the blob stars by observing a more metal-poor sample. Lastly, we observe differences in the abundance trends for a few elements in the optical vs IR. A more detailed investigation on these differences is recommended, especially at such lower [Fe/H], similar to the APOGEE comparison to independent analyses from \citet{jonsson18}.

The synergy of using large spectroscopic survey data from APOGEE and high-resolution follow up spectra like in this work provides a larger discovery space for studying the origins of stars, even after being cannibalized and fully phase-mixed with our Galaxy. The future is especially brighter with the next generation of large spectroscopic surveys (e.g. SDSS-V's Milky Way Mapper, 4MOST, WEAVE) as well as thirty-meter class telescopes (e.g. GMT) for following-up on even fainter and farther ex-situ stars. 
Finding ways to complement these upcoming data sets will not only give a wider view of the Galaxy in volume, but also on its total chemical space. 

\section*{Acknowledgements}
The authors thank the referee for the suggestions that ultimately improved the analysis and the manuscript. AC \& KH have been partially supported by a TDA/Scialog  (2018-2020) grant funded by the Research Corporation and a TDA/Scialog grant (2019-2021) funded by the Heising-Simons Foundation. AC thanks the Large Synoptic Survey Telescope Corporation (LSSTC) Data Science Fellowship Program, which is funded by LSSTC, NSF Cybertraining Grant 1829740, the Brinson Foundation, and the Moore Foundation. AC \& KH acknowledges support from the National Science Foundation grant AST-1907417 and AST-2108736 and from the Wootton Center for Astrophysical Plasma Properties funded under the United States Department of Energy collaborative agreement DE-NA0003843. This work was performed in part at Aspen Center for Physics, which is supported by National Science Foundation grant PHY-1607611.
PJ acknowledges partial support from
FONDECYT REGULAR 1200703 and FONDECYT Iniciaci\'on 11170174.
PD acknowledges funding from the UK Research
and Innovation council (grant number MR/S032223/1).

This paper includes data taken at The McDonald Observatory of The University of Texas at Austin. We thank the staff at McDonald Observatory  for making this project possible.  
Funding for the Sloan Digital Sky Survey IV has been provided
by the Alfred P. Sloan Foundation, the U.S. Department of Energy Office of Science,
and the Participating Institutions. SDSS acknowledges support and resources from the
Center for High-Performance Computing at the University of Utah. The SDSS web site
is \url{www.sdss.org}.
SDSS is managed by the Astrophysical Research Consortium for the Participating
Institutions of the SDSS Collaboration including the Brazilian Participation Group, the Carnegie Institution for Science, Carnegie Mellon University, the Chilean Participation Group, the French Participation Group, Harvard-Smithsonian Center for Astrophysics, Instituto de Astrofasica de Canarias, The Johns Hopkins University, Kavli Institute for the Physics and Mathematics of the Universe (IPMU) / University of Tokyo,
Lawrence Berkeley National Laboratory, Leibniz Institut f\"ur Astrophysik Potsdam
(AIP), Max-Planck-Institut f\"ur Astronomie (MPIA Heidelberg), Max-Planck-Institut f\"ur
Astrophysik (MPA Garching), Max-Planck-Institut f\"ur Extraterrestrische Physik (MPE),
National Astronomical Observatory of China, New Mexico State University, New York
University, University of Notre Dame, Observat\'orio Nacional / MCTI, The Ohio State University, Pennsylvania State University, Shanghai Astronomical Observatory, United
Kingdom Participation Group, Universidad Nacional Aut\'onoma de M\'exico, University
of Arizona, University of colourado Boulder, University of Oxford, University of
Portsmouth, University of Utah, University of Virginia, University of Washington, University of Wisconsin, Vanderbilt University, and Yale University.

\section{Data Availability}
Data is available upon request from the corresponding author. The APOGE DR16 data used in this work can found at \url{https://www.sdss.org/dr16}.

\bibliographystyle{mnras}
\bibliography{references} 




\appendix
\section{Line Selection}
\label{sec:lineselection}

\begin{table*}
\center
\begin{tabular}{cccccc}
\hline \hline
APOGEE ID & Element & Wavelength & log(\textit{gf}) & $\chi$ & \logeps \\
 &  & (\AA) &  & (eV) & (dex) \\
\hline
2M09121759+4408563 & O I & 6363.8 & -10.19 & 0.020 & 7.939 \\
2M09121759+4408563 & Na I & 5688.2 & -0.404 & 2.104 & 4.659 \\
2M09121759+4408563 & Na I & 6160.8 & -1.246 & 2.104 & 4.842 \\
2M09121759+4408563 & Mg I & 5771.1 & -1.724 & 4.346 & 6.460  \\
2M09121759+4408563 & Si I & 5665.6 & -1.940 & 4.920 & 6.226 \\
2M09121759+4408563 & Si I & 5690.4 & -1.773 & 4.930 & 6.173 \\
2M09121759+4408563 & Si I & 5701.1 & -1.773 & 4.930 & 6.332 \\
2M09121759+4408563 & Si I & 5772.1 & -1.653 & 5.082 & 6.313 \\
2M09121759+4408563 & Si I & 5948.5 & -1.130 & 5.082 & 6.342\\
2M09121759+4408563 & Ca I & 5260.4 & -1.179 & 2.521 & 5.171 \\
2M09121759+4408563 & Ca I & 5261.7 & -0.579 & 2.521 & 5.081 \\
... & ... & ... & ... & ... & ...\\
\hline \hline
\end{tabular}
\caption{A sample from the line selection with the following columns: (1) APOGEE ID, (2) Element (3) Wavelength in \AA, (4) log (\textit{gf}), (5) excitation potential, $\chi$, in eV (5), and abundance, \logeps, in dex before scaling to the solar abundance. }
\label{tab:lineselection}
\end{table*}

Here, we discuss the reliability of abundances along with the lines we used to derive them from our optical spectra. A sample of the line selection is listed in Table \ref{tab:lineselection}. For each element, if the majority of stars have more than one line available for abundance determination, we highlight a target that has representative errors in its [X/Fe] i.e., closest to the mean [X/Fe] error values for the whole sample, and examine the range in abundances derived from its individual lines to quantify internal errors. For consistency with the comparison studies, we report the abundances derived with local thermodynamic equilibrium (LTE) but note the difference when non-LTE (NLTE) effects are considered. We discuss the NLTE effects on the abundances further in  Appendix \ref{sec:nlte}. 
\\
\\
\noindent \textbf{O:} We use only the 6363.8~\AA~forbidden line for O which is present in all our spectra. We do not include the O triplet because half of our spectra have poorer quality in that region. The NLTE corrections for 6363.8~\AA~are either zero or unreliable from \citet{sitnova13}, in line with what \citet{amarsi16} found. \citet{heiter2021atomic}~also flag this line as possibly blended, which could alter the measured abundance. There is a known, increasing difference in [O/Fe] between the optical and IR at [Fe/H]$<$1 dex, and the APOGEE [O/Fe] has been noted to not reach as high a value compared to reference works \citep{jonsson18}. 
\\
\\
\noindent \textbf{Na:} We use three neutral lines for Na: 4982.8~\AA, 5688.2 ~\AA, and 6160.8~\AA. All the lines have transition probabilities with undecided quality \citep{heiter2021atomic}. Nonetheless, we use these lines for Na as we derive reasonably small [Na/Fe] scatter in the optical (0.10 dex) compared to the IR (0.35 dex). All three lines are in the target 2M14530873+1611041 that has an APOGEE [Na/Fe] = 0.17 $\pm$ 0.08 dex and [Na/Fe] = -0.11 $\pm$ 0.03 dex from the optical. Its abundance offset in the optical and IR (-0.28 dex) is also representative of the whole sample offset (-0.24 dex). A visual inspection of these lines show that the 5688.2~\AA~line has \logeps~higher by 0.05 dex from the mean value, likely due to the continuum normalization that is affected by a nearby NdII line.  
\\
\\
\noindent \textbf{Mg:} We use the lines 4571.1~\AA~and 5711.1~\AA~with and without NLTE corrections (see Appendix \ref{sec:nlte}). The 4571.1~\AA~line is at times saturated but is caught by BACCHUS in its automatic flagging routine and was only deemed reliable for 9 of the stars. The 5711.1~\AA~line on the other hand has a log$gf$ marked \textit{U} by the GES line list team, meaning the quality is undecided \citep{heiter2021atomic}. We nonetheless use the [Mg/Fe] from the optical because even with the majority of the abundances being determined from only 5711.1~\AA, the [Mg/Fe] scatter in the optical is comparable to the scatter in the IR and is only offset by 0.09 dex.  
\\
\\
\noindent \textbf{Si:} We use 5 neutral lines for Si, 5665.6~\AA, 5690.4~\AA, 5701.1~\AA, 5772.1~\AA, 5948.5~\AA, and some of these lines prone to blending (e.g., 5665.6~\AA, and 5772.1~\AA) are likely causing the higher scatter in [Si/Fe] in the optical compared to the IR. With NLTE corrections from \citet{bergemann13}, the mean [Si/Fe] decreases from 0.29 dex to 0.24 dex but the scatter increases from 0.09 dex to 0.11 dex, meanwhile, the scatter in the IR is lower at 0.05 dex. We therefore opt to use the APOGEE [Si/Fe] for the rest of this work. We also examine the individual line abundances for the target 2M19111765+4359072 which has [Si/Fe] = 0.27 $\pm$ 0.02 dex from APOGEE and [Si/Fe] = 0.30 $\pm$ 0.06 dex from the optical, which has representative errors as the whole sample. Only three out of the maximum five lines were available for this star e.g., 5665.6~\AA, 5701.1~\AA, 5772.1~\AA~and as expected, the lines with the blending flag introduce the scatter: the 5665.6~\AA~line has \logeps~lower by 0.133 dex and the 5772.1~\AA~line has \logeps~higher by 0.105 dex compared to the mean \logeps. 
\\
\\
\noindent \textbf{Ca:} The Ca abundances were derived from 14 lines: 5260.4~\AA, 5261.7~\AA, 5349.5~\AA, 5513~\AA, 5588.7~\AA, 5590.1~\AA, 5857.5~\AA, 6166.4~\AA, 
6169~\AA, 6169.6~\AA, 6455.6~\AA, 6471.7~\AA, 6493.8~\AA, 6499.6~\AA. We note that there are no NLTE corrections for Ca in the spherical MARCS models. The quality of the transition probabilities and the lines are all superb for the Ca lines being considered here aside from 5349.5~\AA, 5513~\AA, and 5857.5~\AA~which are sometimes blended. We examine 2M16153915+4712530 with [Ca/Fe] = 0.21 $\pm$ 0.04 dex from APOGEE and [Ca/Fe] = 0.23 $\pm$ 0.03 dex from our analysis. We find that there is a slight increasing trend for the \logeps~with wavelength, having a \logeps~offset of -0.097 dex from the 5260~\AA~line and +0.155 dex from the 6471.7~\AA~line compared to the mean \logeps.  
\\
\\
\noindent \textbf{Sc:} We consider three singly-ionized Sc lines, 5667.2~\AA, 5684.2~\AA, and 6604.6~\AA, for abundance determination, all of which have accurate transition probabilities. On the other hand, the 5684.2~\AA~and 6604.6~\AA~lines are sometimes blended \citep{heiter2021atomic}. We examine these lines in 2M09121759+4408563 with [Sc/Fe] = 0.09 $\pm$ 0.026 dex, which is representative of the mean [Sc/Fe] abundance and error for the whole sample. The individual \logeps~for 5667.2~\AA, 5684.2~\AA, and 6604.6~\AA~are 1.938 dex, 1.918 dex, and 1.835 dex, respectively, spanning a narrow, $\sim$0.10 dex range in \logeps.
\\
\\
\noindent \textbf{V:} We use 5 neutral V lines, 4875.5~\AA, 5627.6~\AA, 5670.9~\AA, 5727~\AA, and 6111.6~\AA~to derive its abundance. All the lines have accurate transition probabilities and only 5670.9~\AA~and 6111.6~\AA~have fully unblended lines. V has HFS data from \citet{childs79,unkel89,elkashef92,palmeri95,cochrane98,lefebvre02} that are included in the line list \citep{heiter2021atomic}. We examine 2M15035120+2235309 which has an APOGEE [V/Fe] = 0.28 $\pm$ 0.10 dex and [V/Fe] = 0.11 $\pm$ 0.05 dex from this study; this is representative of the mean optical-IR offset and mean [V/Fe] error across the whole sample. The abundance was measured from only four lines in this star, namely 4875.5~\AA, 5670.9~\AA, 5727~\AA, and 6111.6~\AA. Of these lines, 4875.5~\AA~exhibits the maximum positive offset of 0.104 dex and 5670.9~\AA~the maximum negative offset of 0.120 dex from the mean \logeps. 
\\
\\
\noindent \textbf{Cr:} We derived the Cr abundance from a maximum of 9 neutral Cr lines  4789.3~\AA, 4936.3~\AA, 4964.9~\AA, 5247.6~\AA, 5296.7~\AA, 5300.7~\AA, 5345.8~\AA, 5348.3~\AA, 5409.8~\AA~all of which have NLTE corrections aside from 5300.7~\AA~from \citet{bergemann10}. All lines have reliable transition probabilities but only 4936.3~\AA, 5296.7~\AA, and 5348.3~\AA~are unblended. Using just these lines give a similar scatter of 0.08 dex but increases the mean abundance from -0.12 dex to -0.11 dex. We examine the star 2M11193441+4957223 which has a [Cr/Fe] = -0.15 $\pm$ 0.12 dex from APOGEE and -0.16 $\pm$ 0.03 dex from our analysis. All 9 lines are present for this star. There is a slight decreasing trend in \logeps~with wavelength, where the \logeps~from the 4936.3~\AA~line is higher by 0.190 dex compared to the mean \logeps~while the \logeps~from the 5409.8~\AA~line is lower by 0.305 dex. Upon inspection, the 4936.3~\AA~and 5409.8 ~\AA~lines are on the extreme ends of the equivalent widths range at 10.4~m\AA~and 106~m\AA, respectively. 
\\
\\
\noindent \textbf{Mn:} We used a maximum of 6 neutral lines for Mn namely 4783.4~\AA, 5394.7~\AA, 5420.3~\AA, 5432.5~\AA, 6013.5~\AA, 6021.8~\AA~which all have NLTE corrections from \citet{bergemann08}. All the lines considered have reliable transition probabilities but only 5420.3~\AA~and 6021.8~\AA~are unblended.  Mn has HFS data from \citet{handrich69,davis71,luc72,demb79,johann81,brodzinski87,basar03,lefebvre03,bw05} that are included in the line list \citep{heiter2021atomic}. We examine the individual line abundances from the target 2M01575297-0316508 which has [Mn/Fe] = -0.15 $\pm$ 0.09 dex from APOGEE and [Mn/Fe] = -0.14 $\pm$ 0.04 dex from our analysis. The [Mn/Fe] for this star was derived from the lines  5420.3~\AA, 5432.5~\AA, 6013.5~\AA, and 6021.8~\AA~where the \logeps~from the 5420.3 ~\AA~line is higher by 0.089 dex, and from the 5432.5~\AA~line lower by 0.085 dex compared to the mean \logeps; both are strong lines with equivalent widths $>$ 110~m\AA.  
\\
\\
\noindent \textbf{Co:} We use a maximum of 7 neutral Co lines, 5212.7~\AA, 5331.40~\AA, 5369.6~\AA, 5530.8~\AA, 5647.2~\AA, 5915.5~\AA, 6771~\AA, to derive Co abundance, all of which have NLTE corrections from \citet{bergemann10b}. All lines have accurate transition probabilities but only 5331.40~\AA, 5647.2~\AA, and 6771~\AA~are unblended. The whole sample has mean [Co/Fe] = 0.22 dex with scatter 0.11 dex, and the NLTE corrections increase both mean and scatter to 0.52 $\pm$ 0.23 dex. 
Co has HFS data from \citet{pickering96} that are included in the line list \citep{heiter2021atomic}. We examined the individual line abundances for the target 2M13141390+1817489 which has [Co/Fe] = -0.26 $\pm$ 0.06 dex from APOGEE and [Co/Fe] = 0.16 $\pm$ 0.04 dex from this study, which is representative of the optical-IR offset and the mean error value for the whole sample. For this star, the 5212.7~\AA~line exhibits a -0.148 dex offset while the 6771~\AA~line a +0.115 dex offset from the mean \logeps~value. Upon inspection, the continuum region around the 5212.7~\AA~line and the near-saturation of the 6771~\AA~line appear to affect the measured abundance. 
\\
\\
\noindent \textbf{Ni:} We use a maximum of 10 neutral Ni lines to derive its abundance, namely 4873.4~\AA, 5003.7~\AA, 5035.4~\AA, 5137.1~\AA, 5435.9~\AA, 5587.8~\AA, 6176.8~\AA, 6482.8~\AA, 6643.6~\AA, and 6767.8~\AA. With this line selection, we derive a mean [Ni/Fe] = -0.03 dex with a scatter of 0.07 dex for the blob stars. All the lines have reliable transition probabilities aside from 6482.8~\AA~where the quality of the data is undecided \citep{heiter2021atomic}. The lines 4873.4~\AA, 5003.7~\AA, 5035.4~\AA, 5137.1~\AA, and 5587.8~\AA~are sometimes blended. Considering just the lines of highest quality (i.e., 5435.9~\AA, 6176.8~\AA, 6643.6~\AA, and 6767.8~\AA), the modified mean abundance is 0.03 $\pm$ 0.06 dex. We examined the individual line abundances for the target 2M15410952+3014067 which has [Ni/Fe] = 0.05 $\pm$ 0.04 from APOGEE and [Ni/Fe] = 0.01 $\pm$ 0.05 from this study. There is a slight increasing \logeps~with wavelength, where the line at 5035.4~\AA~gives a \logeps~that is 0.270 dex below the mean \logeps~while the line at 6643.6~\AA~gives a \logeps~that is 0.257 dex above the mean value. Upon inspection, the 6643.6~\AA~line is near saturation and has an equivalent width of 151~m\AA, likely causing the overestimation in abundance. 
\\
\\
\noindent \textbf{Cu:} We use a maximum of 4 neutral Cu lines, 5105.5~\AA, 5218.2~\AA, and 5700.2~\AA, 5782.1~\AA, in deriving its abundance. All lines considered have accurate transition probabilities but are flagged as prone to blending \citep{heiter2021atomic}.  HFS data for Cu were taken from \citet{fischer67,bergstrom89,hermann93}. We analyze the individual line abundances for 2M16445645+4230263 which has [Cu/Fe] = 0.21 $\pm$ 0.05 dex from APOGEE and [Cu/Fe] = -0.44 $\pm$ 0.05 dex from this study, which is representative of the total sample with respect to the errors and the optical-to-IR abundance offset. Although the measured [Cu/Fe] from the optical and IR are starkly different for this star, it is actually representative of the overall offset  as well as the mean error in [Cu/Fe] for the entire sample. The 5700.2~\AA~line has \logeps~lower by 0.117 dex than the mean while the 5782.1~\AA~line has \logeps~higher by -0.107 dex, and both lines are reliable upon visual inspection. 
\\
\\
\noindent \textbf{Zn:} We only use one neutral Zn line at 4810.5~\AA, which has reliable transition probability but is marked as blended \citep{heiter2021atomic} which drives the scatter.
\\
\\
\noindent \textbf{Y:} We used 4 singly-ionized Y lines, 5087.4~\AA, 5200.4~\AA, 5205.7~\AA, and 5289.8~\AA~to derive [Y/Fe]. All the lines considered have accurate transition probabilities but 5200.4~\AA~and 5205.7~\AA~are possibly prone to blending. We inspect the individual line abundances for the target 2M09381836+3706176 ([Y/Fe] = 0.05 $\pm$ 0.04 dex) and find that the \logeps~from the 5289.8~\AA~line generally increases the total abundance, and is 0.142 dex higher than the mean \logeps. This line also has the lowest equivalent width (i.e., 16.3~m\AA) of the lines available. 
\\
\\
\noindent \textbf{Zr:} We use 4 singly-ionized Zr line, 4317.3~\AA, 6127.4~\AA, 6134.6~\AA, and 6143.2~\AA~in the abundance determination. The 4317.3~\AA~line can sometimes be saturated, but all the other lines are unblended and have reliable transition probabilities \citep{heiter2021atomic}. We inspect the individual line abundances for the target 2M16153915+4712530 which has [Zr/Fe] = 0.19 $\pm$ 0.10 dex from our analysis, representative of the mean [Zr/Fe] abundance and error for our sample. The 4317.3~\AA~line has a \logeps~higher by 0.125 dex compared to the mean \logeps~which upon inspection, we realize is due to the continuum normalization in the spectral window. 
\\
\\
\noindent \textbf{Ba:} We derive Ba abundance from only one singly-ionized line at 5853.7~\AA~which is deemed reliable from \citep{heiter2021atomic}. HFS data for Ba were taken from \citet{becker81,silverans86,villamoes93} for the Gaia-ESO line list. 
\\
\\
\noindent \textbf{La:} We used the singly-ionized La lines 4322.5~\AA, 4333.8~\AA, 4804~\AA, 5303.5~\AA, 5805.8~\AA, and 6390.5~\AA~to determine La abundance. All the lines considered could be prone to blending through our visual inspection and from flags determined from \citep{heiter2021atomic}. This blending as well as the continuum normalization, especially in the bluer part of the spectrum, can skew the abundance as we have examined for the star 2M16064562-2247445 ([La/Fe] = 0.21 $\pm$ 0.04 dex), where the 4322.5~\AA~line is 0.183 dex lower and the 4333.8 ~\AA~line is 0.214 dex higher than the mean \logeps. 
\\
\\
\noindent \textbf{Ce:} We use only one line to determine Ce abundance, 5247.2~\AA, which is flagged as reliable \citep{heiter2021atomic}. This gives a mean [Ce/Fe] value of 0.31 dex $\pm$ 0.22, which is higher compared to the mean [Ce/Fe] from APOGEE (-0.04 dex $\pm$ 0.32). Taking into account that our Ce solar abundance is higher by 0.31 dex compared to \citet{asplund09} would make our derived [Ce/Fe] agree with the literature. However, for consistency, especially in deriving the neutron-capture element-to-element abundance ratios, we use the solar abundance that we derived independently. 
\\
\\
\noindent \textbf{Nd:} We obtain Nd abundances from a maximum of 4 singly-ionized Nd lines at 4446.4~\AA, 4914.4~\AA, 5092.8~\AA, and 5319.8~\AA. The line at 4446.4~\AA~can be affected by blending due to a nearby FeII line. All the lines considered have reliable transition probabilities but are flagged to be possibly blended and are prone to HFS \citep{heiter2021atomic}. HFS data for Nd were taken from \citet{ma04,rosner05}. We inspected the target 2M19104370-6001040 wherein all 4 Nd lines are present and give [Nd/Fe] = 0.49 $\pm$ 0.03 dex. The 5092.8~\AA~line abundance is higher by 0.093 dex while the 5319.8~\AA~line abundance is lower by 0.135 dex compared to the mean \logeps. Upon inspection, the fit to the 5319.8~\AA~line, though visually is a good fit and is also flagged by BACCHUS as good, does have the highest $\chi^2$ among the lines available.
\\
\\
\noindent \textbf{Eu:} We consider the singly-ionized Eu line at 6645.1~\AA~for its abundance determination. From \citep{heiter2021atomic}, this line has reliable and accurate transition probability but could be prone to blending. HFS is accounted for this line using data from \citet{villemoes92}.

\section{NLTE corrections}
\label{sec:nlte}
\begin{figure}
\centering
\includegraphics[width=0.4\textwidth]{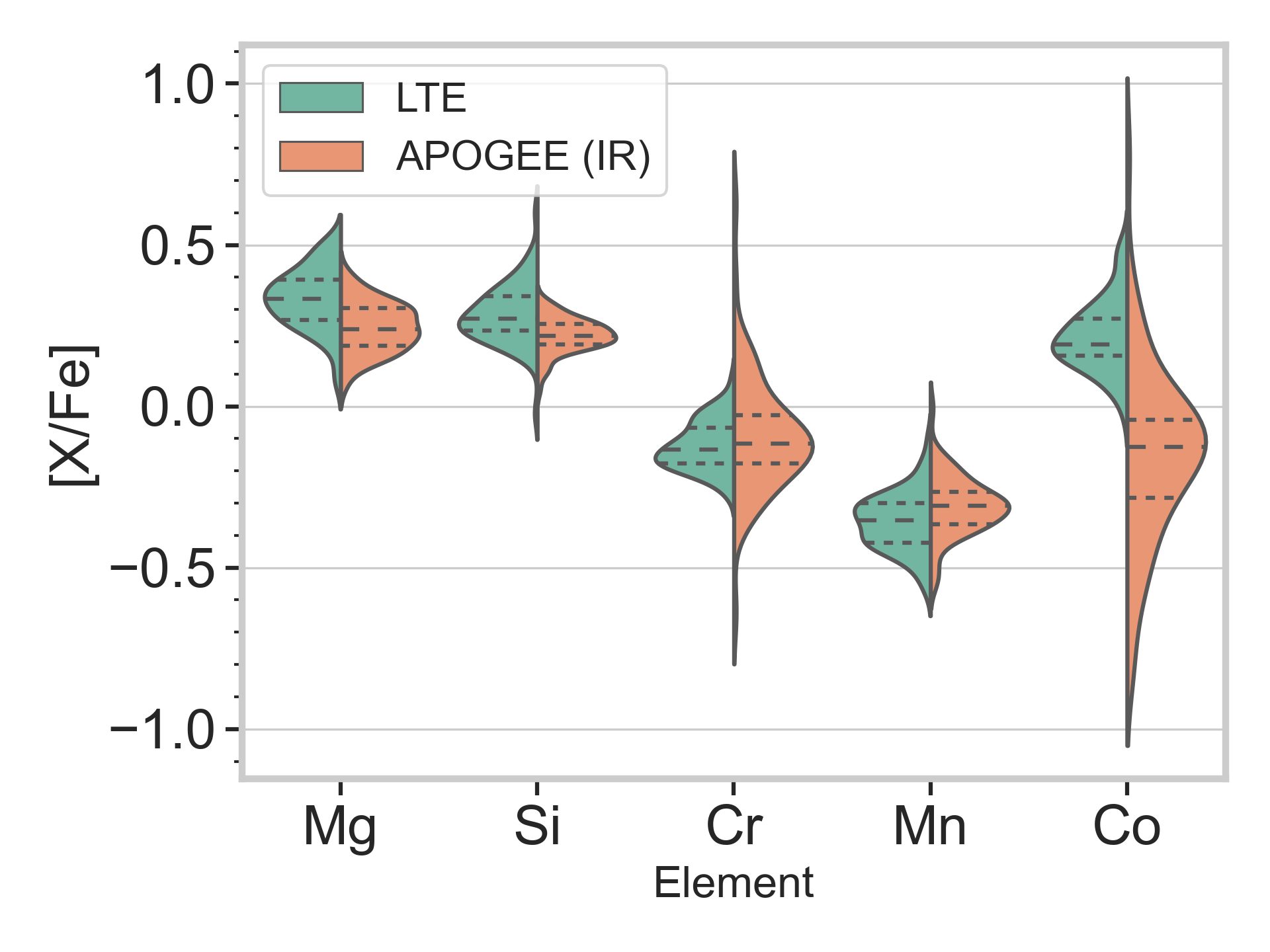}
\includegraphics[width=0.4\textwidth]{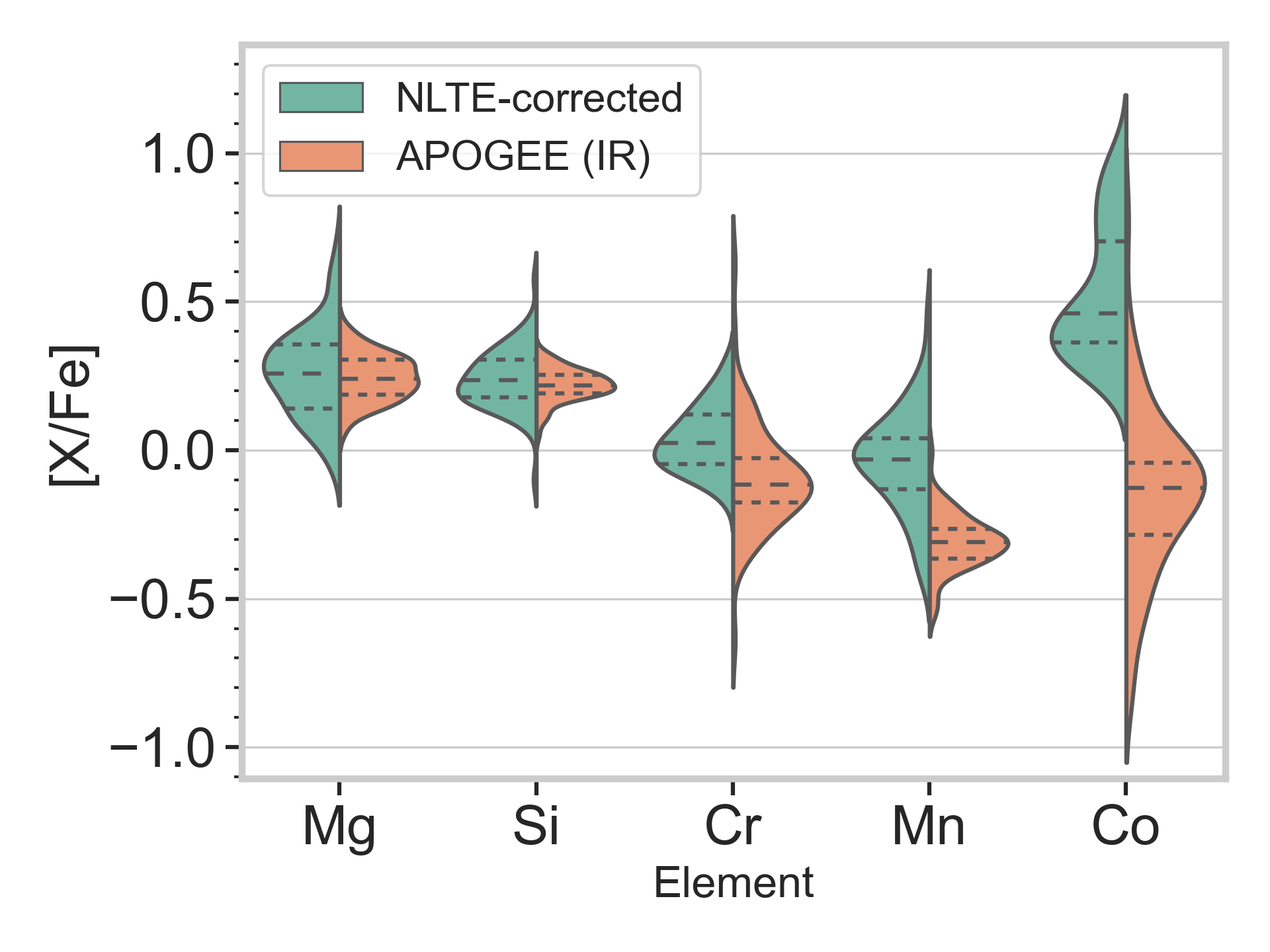}
\caption{\textit{\textbf{Violin plot comparing [X/Fe] abundances with and without NLTE corrections}}. The top plot is directly taken from Figure \ref{fig:scatter}  but only with the elements that we have NLTE corrections for, shown in the bottom plot. The $\alpha$ elements Mg and Si have better accuracy compared to APOGEE with the NLTE corrections, but Cr, Mn, and Co increase both in offset and scatter compared to their APOGEE counterparts. } 
\label{fig:violin_nlte}
\end{figure}

\begin{figure*}
\centering
\includegraphics[width=0.9\textwidth]{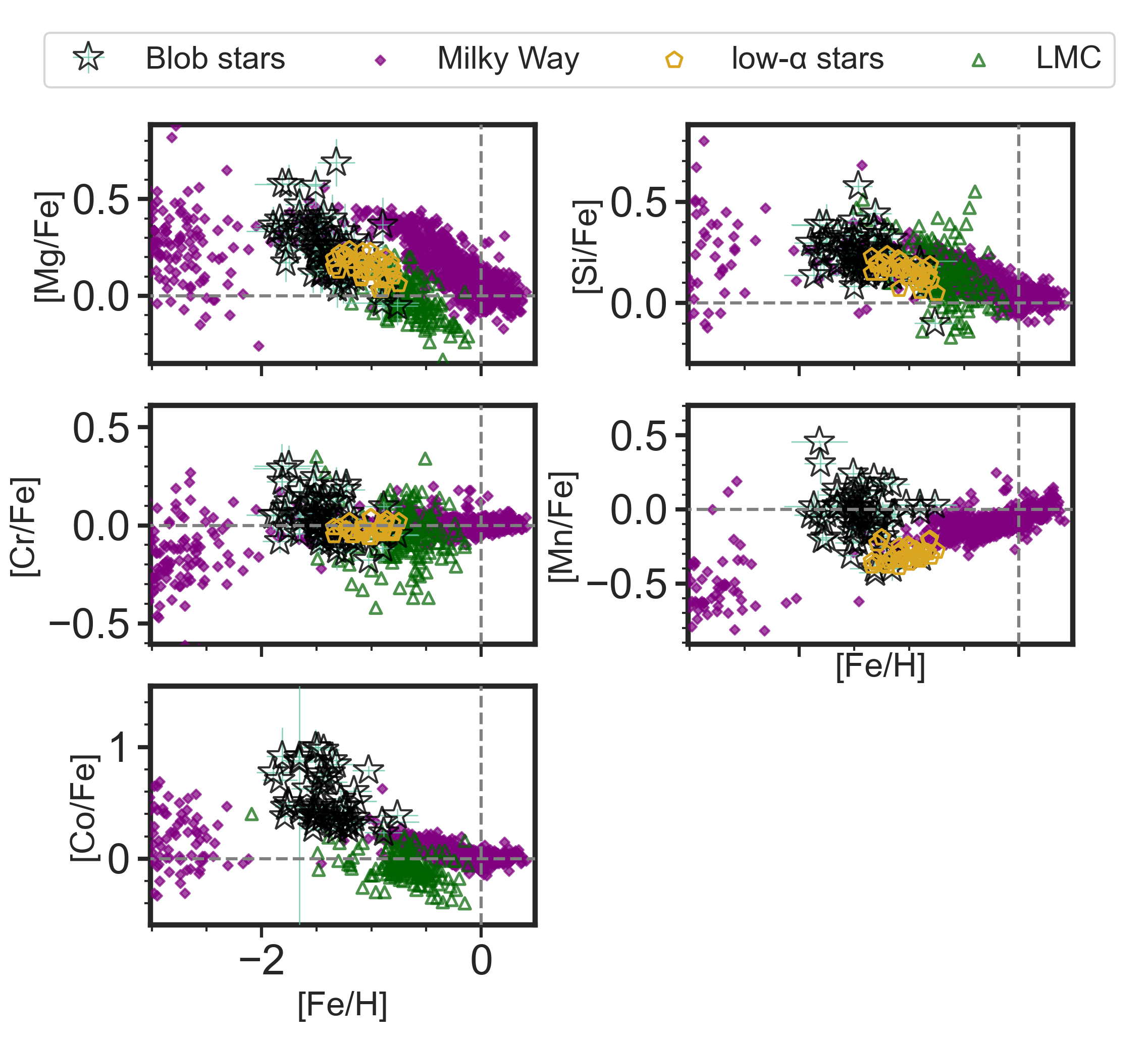}
\caption{\textit{\textbf{[X/Fe] vs [Fe/H] with NLTE corrections}}. Legends are similar to Figures \ref{fig:alpha},\ref{fig:lightodd},\ref{fig:ironpeak}, and \ref{fig:ncapture}. These show more clearly that there is better agreement between the low-$\alpha$ stars and our sample in their Mg and Si abundances, but that the scatters and offsets increase at the lower-metallicity end for Cr and Mn, and at all metallicities for Co, compared to their LTE abundances. } 
\label{fig:xfe_nlte}
\end{figure*}

We also applied NLTE corrections for the elements Mg, Si, Cr, Mn, and Co and contrast their trends with [Fe/H] with the LTE case. Corrections were obtained from \citet{bergemann17} for Mg, \citet{bergemann13} for Si, \citet{bergemann10} for Cr, \citet{bergemann08} for Mn, and \citet{bergemann10b} for Co through \url{https://nlte.mpia.de/gui-siuAC_secE.php} using spherical 1D MARCS models as is apt for our sample of red giant branch stars. The NLTE corrections for the 6363.8 ~\AA~O line is zero or unreliable while there are no Ca NLTE corrections in the MARCS models. Across all these elements, the NLTE corrections increase the scatter in the individual abundance distributions, more so for Cr, Mn, and Co as is shown in Figure \ref{fig:violin_nlte} and listed in Table \ref{tab:stats_nlte}. This is seen more clearly in the individual [X/Fe] vs [Fe/H] for these elements shown in Figure \ref{fig:xfe_nlte}. Specifically, this discrepancy increases at the lower-metallicity end of the sample. We list the stars and the available NLTE corrections in Table \ref{tab:nlte_cor}.

The NLTE corrections had the following effects on the abundance trends: for [Mg/Fe], the mean decreased to 0.26 dex from 0.33 dex but the scatter increased to 0.15 dex from 0.09 dex; for [Cr/Fe], the mean increased to positive values i.e., 0.04 dex and the scatter increased to 0.11 dex; for [Mn/Fe], the abundance increased at the lower-metallicity end of the sample, as similarly seen by \citet{battistini15} when they applied Mn NLTE corrections for Milky Way disk stars. With the NLTE corrections, the mean [Mn/Fe] (-0.04 dex) is higher compared to the LTE case (-0.13 dex), and the scatter more than doubles to 0.19 dex compared to 0.08 dex at LTE.

\begin{table}
\begin{center}
\begin{tabular}{c|ccc|ccc|cc}
\hline \hline
(1) & (2) & (3) \\
$\rm [X/Fe]$ & $\rm \mu_{optical}$ (dex) & $\rm \sigma_{optical}$  (dex) \\ 
\hline
Mg & 0.26 & 0.15 \\
Si & 0.24 & 0.11 \\
Cr & 0.04 & 0.11 \\
Mn & -0.04 & 0.17 \\
Co & 0.53 & 0.22 \\
\hline \hline
\end{tabular}
\caption{Summary statistics for the [X/Fe] trends with NLTE corrections. }
\label{tab:stats_nlte}
\end{center}
\end{table}

\begin{table*}
\begin{tabular}{c|c|c|c|c|c|c|c|c|c|c}
\hline
\hline
star & 6363.850 & 4571.090 & 5711.070 & 5665.600 & 5690.470 & 5701.140 & 5708.440 & 5772.150 & 5948.541 & ... \\
& O & Mg & Mg & Si & Si & Si & Si & Si & Si & ...\\
\hline
2M09121759+4408563 & 0.0 & 0.33 & -0.033 & -0.094 & -0.1 & -0.091 & -0.104 & -0.035 & -0.042 & ...\\
2M11115726+4551087 & 0.0 & 0.367 & -0.152 & -0.051 & -0.058 & -0.049 & -0.068 & -0.023 & -0.031 & ...\\
2M12071560+4622126 & 0.0 & 0.34 & -0.082 & -0.031 & -0.035 & -0.031 & -0.037 & -0.014 & -0.017 & ...\\
2M13581572+2602122 & 0.0 & 0.242 & 0.003 & -0.101 & -0.101 & -0.096 & -0.097 & -0.033 & -0.037 & ...\\
2M09381836+3706176 & 0.0 & 0.142 & -0.177 & -0.032 & -0.036 & -0.031 & -0.042 & -0.015 & -0.02 & ...\\
2M11482205-0030318 & 0.0 & 0.346 & -0.085 & -0.059 & -0.061 & -0.057 & -0.061 & -0.021 & -0.026 & ...\\
2M10013420+4345558 & 0.0 & 0.366 & -0.132 & -0.044 & -0.049 & -0.043 & -0.055 & -0.02 & -0.026 & ...\\
2M13015242+2911180 & 0.0 & 0.35 & -0.033 & -0.089 & -0.096 & -0.088 & -0.099 & -0.034 & -0.041 & ...\\
2M14181562+4651580 & 0.0 & 0.332 & -0.01 & -0.102 & -0.108 & -0.099 & -0.114 & -0.038 & -0.046 & ...\\
2M15410952+3014067 & 0.0 & 0.226 & -0.098 & -0.028 & -0.031 & -0.028 & -0.033 & -0.012 & -0.015 & ...\\
... & ... & ... & ... & ... & ... & ... & ... & ... & ... & ...\\
\hline
\hline
\end{tabular}
\caption{Snippet of the lines with NLTE corrections (columns) for each target (rows). The full table including the NLTE corrections for the rest of the targets and for Cr, Mn, and Co will be available online.}
\label{tab:nlte_cor}
\end{table*}

\bsp	
\label{lastpage}
\end{document}